\documentclass[amsmath,amssymb,aps,prx,notitlepage,superscriptaddress, longbibliography, reprint]{revtex4-2}
\usepackage{graphicx}
\usepackage{dcolumn}
\usepackage{bm}
\usepackage{braket}
\usepackage{amsthm}
\usepackage{epstopdf}
\usepackage{amssymb}
\usepackage{amsmath}
\usepackage{bbold}
\usepackage{hyperref}
\usepackage{comment}
\usepackage{xcolor}
\usepackage[normalem]{ulem}
\usepackage[utf8]{inputenc}

\theoremstyle{definition}

\begin{document}
\preprint{APS/123-QED}
\title{Classical simulation of lossy boson sampling using matrix product operators}
\author{Changhun Oh}%
\affiliation{Pritzker School of Molecular Engineering, University of Chicago, Chicago, Illinois 60637, USA}
\author{Kyungjoo Noh}
\affiliation{AWS Center for Quantum Computing, Pasadena, California 91125, USA}
\thanks{This was work done before K.N. joined AWS Center for Quantum Computing.}
\author{Bill Fefferman}
\affiliation{Department of Computer Science, University of Chicago, Chicago, Illinois 60637, USA}
\author{Liang Jiang}
\affiliation{Pritzker School of Molecular Engineering, University of Chicago, Chicago, Illinois 60637, USA}
\affiliation{AWS Center for Quantum Computing, Pasadena, California 91125, USA}
\date{\today}
\begin{abstract}
Characterizing the computational advantage from noisy intermediate-scale quantum (NISQ) devices is an important task from theoretical and practical perspectives.
Here, we numerically investigate the computational power of NISQ devices focusing on boson sampling, one of the well-known promising problems which can exhibit quantum supremacy.
We study hardness of lossy boson sampling using matrix product operator (MPO) simulation to address the effect of photon-loss on classical simulability using MPO entanglement entropy (EE), which characterizes a running time of an MPO algorithm.
An advantage of MPO simulation over other classical algorithms proposed to date is that its simulation accuracy can be efficiently controlled by increasing an MPO's bond dimension.
Notably, we show by simulating lossy boson sampling using an MPO that as an input photon number grows, its computational cost, or MPO EE, behaves differently depending on a loss-scaling, exhibiting a different feature from that of lossless boson sampling.
Especially when an output photon number scales faster than the square root of an input photon number, our study shows an exponential scaling of time complexity for MPO simulation.
On the contrary, when an output photon number scales slower than the square root of an input photon number, MPO EE may decrease, indicating that an exponential time complexity might not be necessary.
\end{abstract}

\maketitle

\section{Introduction}
Quantum computers are expected to provide a computational advantage that enables solving problems that lie beyond the computational power of classical computers \cite{nielsen2002quantum}.
Ultimately, quantum computers are demanded to be fault-tolerant and scalable to solve various practical problems that no known classical algorithm can efficiently solve such as integer factorization \cite{shor1994algorithms}. However, since fault-tolerant quantum computing is not immediately feasible with current technology, there has been a huge interest in achieving ``quantum supremacy'' with noisy intermediate-scale quantum (NISQ) \cite{preskill2018quantum} devices. In particular, various sampling problems, such as IQP \cite{bremner2011classical}, boson sampling \cite{aaronson2011computational, hamilton2017gaussian}, Fourier sampling \cite{fefferman2016}, and random circuit sampling (RCS) \cite{boixo2018characterizing}, have been proposed as promising candidates for demonstrating quantum advantage over classical computers. Indeed, there are various complexity-theoretic hardness results which show that these problems cannot be tackled efficiently by a classical computer under reasonable conjectures \cite{aaronson2011computational,aaronson2016bosonsampling,aaronson2016complexity,bouland2019complexity,movassagh2018proof,movassagh2019quantum}.  

Recently, RCS was implemented in a state-of-the-art superconducting qubit system comprising $53$ qubits which are connected in a planar architecture via two-qubit gates of error rates lower than $0.6\%$ \cite{arute2019quantum}. Remarkably, it has been estimated that it would take $2.5$ days \cite{pednault2019leveraging}, $20$ days \cite{huang2020classical}, and $10000$ years \cite{arute2019quantum} to solve an equivalent computational task using one of the best available classical supercomputers. While the estimates vary, it has become evident that classical simulation of the state-of-the-art NISQ systems can only be done, if ever possible, using the most powerful supercomputer available today. 


Aside from demonstrating quantum computational advantage, RCS may prove to have practical applications such as certified random number generation \cite{aasronson2020}. Regardless of the usefulness of the sampling problems, the question of whether a classical computer can simulate random circuits of a NISQ device has important implications in the field of quantum computing: by studying classical simulability of noisy versions of sampling problems, we can sharpen our understanding of how noise limits quantum computational power and hence the utility of a NISQ device.


It is worth noting that many classical algorithms for simulating NISQ systems do not take advantage of the fact that NISQ devices are noisy. That is, many classical simulation methods become unavoidably ineffective for simulating large quantum systems (consisting of, e.g., $70$ qubits) due to exponentially large Hilbert space, even if such systems are noisier than a smaller system which can be classically simulated. On the other hand, various efficient simulation methods based on matrix product state (MPS) and matrix product operator (MPO) \cite{vidal2003efficient} have recently been proposed for simulating large but noisy quantum systems \cite{huang2019simulating,napp2020efficient,zhou2020what,noh2020efficient}. These methods take advantage of the fact that noise in quantum circuits limits the growth of quantum entanglement in NISQ devices and thus use MPS or MPO to describe such NISQ systems with bounded entanglement in a compressed manner. Hence, these MPS-based simulation methods allow us to systematically explore the adverse effects of noise on the computational power of a NISQ device.

Among various proposals for quantum supremacy experiments, we study boson sampling, which is one of the promising candidates expected to exhibit quantum supremacy.
Boson sampling has been proven to be classically intractable under plausible assumptions \cite{aaronson2011computational}.
More precisely, there is no classical algorithm under complexity-theoretic conjectures that approximately samples the outcomes of an ideal boson sampling in polynomial time as an input photon number grows.
Thanks to its experimental setup's relatively simple structure, experimental implementations of boson sampling are rapidly developing \cite{broome2013photonic, spring2013boson, tillmann2013experimental, crespi2013integrated, spagnolo2014experimental, carolan2014experimental, carolan2015universal, bentivegna2015experimental, zhong201812, zhong2019experimental, paesani2019generation, he2017time, loredo2017boson, wang2017high, wang2018toward, wang2019boson, zhong2020quantum}.
Remarkably, the most recent boson sampling experiment and Gaussian boson sampling experiment, a variant of boson sampling, have detected up to 14 photon clicks out of 20 single-photon input \cite{wang2019boson} and up to 76 photons from squeezed states with squeezing parameters ranging from 1.34 to 1.84 \cite{zhong2020quantum}, respectively.
Also, there has been a proposal to employ Gaussian boson sampling \cite{hamilton2017gaussian, zhong2020quantum} to generate molecular vibronic spectra, which has recently been experimentally conducted \cite{wang2020efficient} using a superconducting bosonic processor.

The experimental platform of boson sampling is based on linear optics (beam splitters and phase shifters), as well as single photon sources and detectors.
Although these apparatuses are readily available in experiments, current quantum optics experiments still severely suffer from various imperfections such as impurity of single photons, photon-loss in the circuit, and inefficiency of photo-detectors.
Theoretically, the aforementioned imperfections can be simply modeled as photon-loss, and there have been many theoretical studies to address the hardness of lossy boson sampling \cite{aaronson2016bosonsampling, oszmaniec2018classical, garcia2019simulating, renema2018classical, qi2020regimes}.
Particularly, it is proven that when the input photon number is $N$ and only a constant number of photons $n$ is lost in the system so that we detect $N_\text{out}=N-n$ number of photons, the hardness of boson sampling is maintained \cite{aaronson2016bosonsampling}. 
On the other hand, it can be easily shown that if only $N_\text{out}\propto \log_2 N$ number of photons remain in the measurement, an efficient classical simulation is possible \cite{aaronson2016bosonsampling, clifford2018classical}.
Recently, it has been shown that when $N_\text{out}\propto \sqrt{N}$ number of photons survive before measurement \cite{oszmaniec2018classical, garcia2019simulating}, the lossy boson sampling can be efficiently simulated with a constant error.
The basic idea of such algorithms is that an input state of boson sampling after a large amount of loss can be approximated by thermal states or so-called particle-separable states, which can be employed to simulate the boson sampling efficiently as an input photon number grows.

Meanwhile, a limitation of the algorithms presented above is apparent that once a system's parameters are given, the closest thermal state and particle-separable state are determined. 
Thus, the simulation's accuracy is fixed and cannot be improved by using more computational time.
For this reason, the algorithms may not be applicable to an intermediate size of lossy boson sampling where a loss rate is not large enough to approximate an input state by thermal states or particle-separable states accurately.
Another algorithm to simulate lossy boson sampling employs the fact that outcomes of a large degree of multiphoton interference are suppressed by photon-loss, which allows us to control the approximation error by setting a threshold of the degree according to a target error \cite{renema2018classical}.
In this work, we employ a different approach to simulate lossy boson sampling to overcome the limitation of fixed accuracy by using MPOs \cite{verstraete2006matrix, zwolak2004mixed}.
Specifically, an approximation error of MPO simulation can be manipulated to achieve a target error $\epsilon$ in polynomial
time in $1/\epsilon$ \cite{verstraete2006matrix}.
Therefore, MPO simulation enables us to simulate an intermediate size of boson sampling and achieve a tunable accuracy in an efficient way.
In addition, MPO allows us to compute probabilities approximately.


We characterize how computational cost changes as an input photon number grows using the so-called MPS/MPO entanglement entropy (EE) \cite{verstraete2006matrix, schuch2008entropy}.
In fact, MPO has been used to simulate an intermediate size of boson sampling \cite{huang2019simulating}, where lossy boson sampling was simulated for fixed system size with different loss rates and it was numerically shown that boson sampling with a large amount of photon-loss requires only a small amount of computational cost using MPO EE. 
In this paper, using MPO simulation and MPO EE, we demonstrate how the computational cost of a classical simulation changes as the system size varies, namely input photon number, for various loss scalings.

We first investigate lossless boson sampling with an MPS method for a different number of input photons and modes to compare with lossy boson sampling.
We obtain a consistent numerical result with the theoretical hardness result of ideal boson sampling that the maximum MPS EE over all possible bipartitions linearly increases as the number of input photons grows, suggesting that MPS simulations of ideal boson sampling necessitate an exponential time cost.
More importantly, we investigate classical simulability of lossy boson sampling using MPO simulations.
Particularly, we consider a power-law scaling, i.e., $N_\text{out}\propto N^\gamma$ ($0<\gamma\leq 1$).
In this scaling, a simple procedure using binomial sampling of a pure input state followed by the Clifford-Clifford algorithm \cite{clifford2018classical, clifford2020faster}, the fastest known boson sampling algorithm, does not allow an efficient simulation (see Sec. \ref{sec:lossy}).
Our numerical results show that for a constant loss rate, i.e., $\gamma=1$, the MPO simulation requires an exponential computational time in input photon numbers.
We also analytically show that for $\gamma >1/2$, the required computational cost grows exponentially in an asymptotic regime.
Moreover, we show that for some power-law loss-scaling, such as $\gamma=1/4, 1/2$, an MPO EE drops or increases only logarithmically even if the number of output photons increases in the system. 
Such a behavior of MPO EE might allow an efficient classical simulation, while the scaling of computational cost cannot be determined solely by MPO EE in this regime.


Our paper is organized as follows. 
In Sec. \ref{sec:bs} we introduce basic concepts of ideal boson sampling and lossy boson sampling, taking into account photon loss.
In Sec. \ref{sec:method}, we introduce MPS and MPO methods to simulate boson sampling and MPS and MPO EE, which determines the classical simulability from MPS and MPO methods in Sec. \ref{sec:mpoee}.
Using the provided simulation procedure, we show our numerical simulation results in Sec. \ref{sec:results}.
We first demonstrate that MPS simulation for lossless boson sampling is inefficient using MPS in Sec. \ref{sec:lossless}.
We then show different behaviors of simulability in the simulation of lossy boson sampling using MPO depending on the loss-scaling in Sec. \ref{sec:lossy}.
We also show that the simulation errors can be controlled efficiently.
Finally, we summarize our results in Sec. \ref{sec:conclusion}.

\begin{figure*}[t]
\includegraphics[width=500px]{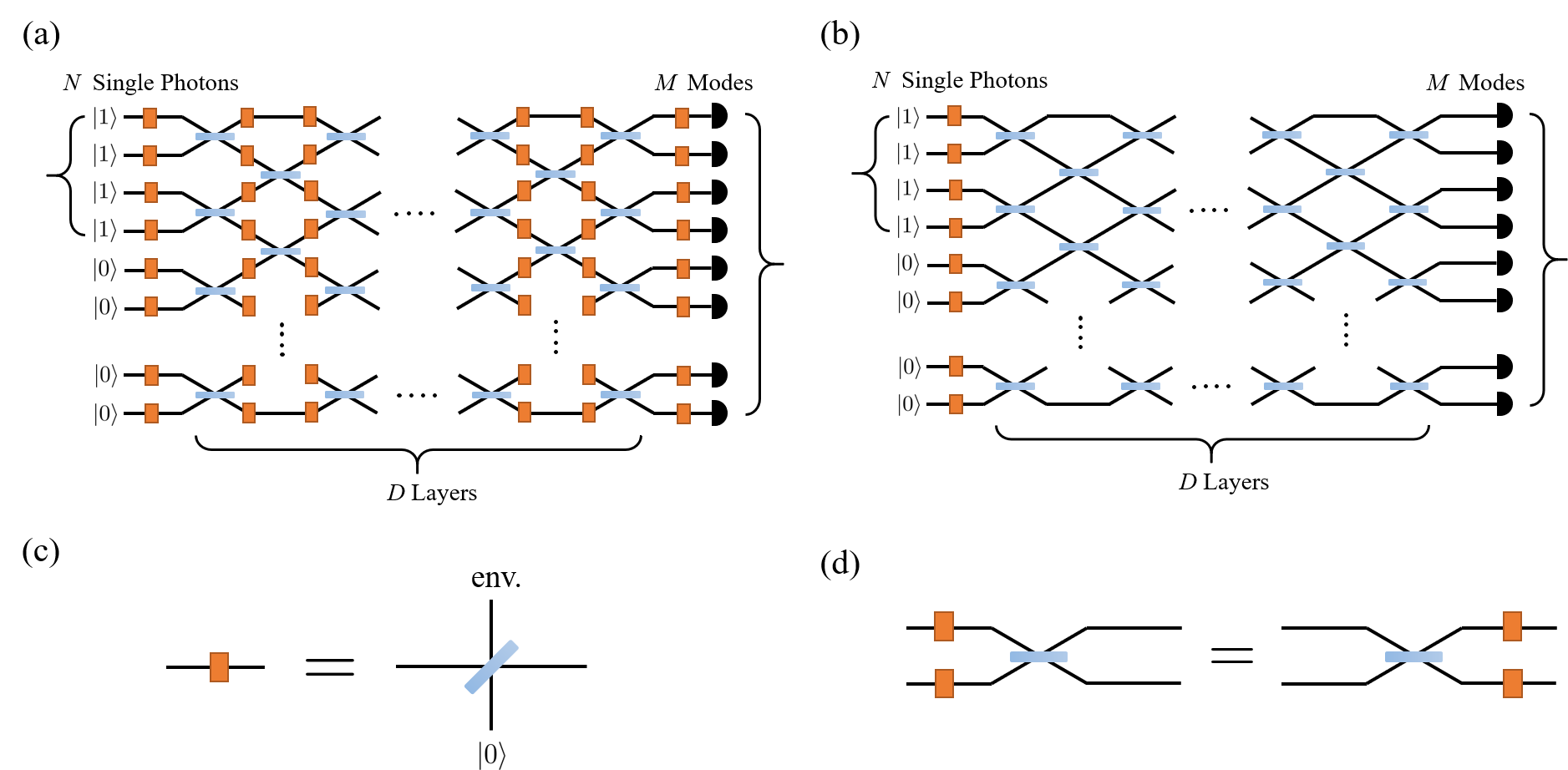}
\caption{Lossy boson sampling circuit. (a) We start with $N$ single photons as an input state and measure the output state by photon-number resolving detectors (or single-photon detectors) after an $M$-mode Haar-random unitary circuit composed of $D$ layers of beam splitters. Imperfection, which are modeled as photon-loss described by (c), occurs in preparation of single photons, beam splitter circuits, and detectors. (b) Assuming a uniform photon-loss rate for different modes, photon-loss channels and beam splitter commute, as shown in (d), so that all the photon-loss can be moved to the preparation step.}
\label{fig:circuit}
\end{figure*}

\section{Lossy Boson Sampling}\label{sec:bs}
Let us consider boson sampling circuits consisting of $D$ layers of beam splitters in $M$ bosonic modes $\{\hat{a}_j\}_{j=1}^M$ with $N$ indistinguishable single photons as an input state $|\psi_\text{in}\rangle=|1\rangle^{N}|0\rangle^{M-N}$.
Each layer of the circuit is composed of beam splitters with a random transmissivity and phase shift as illustrated in Fig.\ \ref{fig:circuit} (a). More explicitly, 
a beam splitter between two adjacent modes $\hat{a}^\dagger$ and $\hat{b}^\dagger$ transforms the modes operators as
\begin{align}
\begin{pmatrix}
\hat{a}^\dagger \\
\hat{b}^\dagger
\end{pmatrix}
\to
\begin{pmatrix}
\cos \theta & -e^{i\phi}\sin\theta  \\ 
e^{-i\phi}\sin \theta & \cos\theta
\end{pmatrix}
\begin{pmatrix}
\hat{a}^\dagger \\
\hat{b}^\dagger
\end{pmatrix},
\end{align}
where $\cos\theta$ and $\sin\theta$ correspond to the transmittance and reflectance of the beam splitter, and $\phi$ is a relative phase shift. After applying $D$ layers of beam-splitter operations, we get a passive unitary circuit $\hat{U}$ which transforms modes operators as
\begin{align}
\hat{a}_j^\dagger\to \sum_{k=1}^M U_{jk}\hat{a}_k^\dagger , 
\end{align}
where $U_{jk}$'s are the matrix elements of an $M\times M$ unitary matrix $U$. We choose the circuit depth $D$, random transmissivities ($\cos\theta$), and phase shifts ($\phi$) such that the resulting unitary matrix $U$ is given by a Haar-random $M\times M$ unitary matrix. In particular, it was shown in Ref.\ \cite{russell2017direct} that a circuit depth $D = M$ suffices to generate an $M\times M$ Haar-random unitary matrix $U$. The transmissivities and phase shifts are chosen randomly following the sampling procedure in Ref.\ \cite{russell2017direct}. See Appendix \ref{appendix:Haar-random} for more details.

After going through all the beam splitters, output modes are measured by photon-number detectors.
Surprisingly, the output probability of the seemingly simple structure of linear optical circuits is hard to compute on average, which is formally written as
\begin{align}
|\langle t_1,\dots, t_M|\hat{U} |s_1,\dots, s_M\rangle|^2=\frac{|\text{Per}(U_{S,T})|^2}{t_1!\cdots t_M! s_1!\cdots s_M!}.
\end{align}
Here, we assumed $|s_1,\dots,s_M\rangle$ an input state and $|t_1,\dots,t_M\rangle$ as an output state with $s_j$ and $t_j$ being the photon number at $j$th mode, and
\begin{align}
\text{Per}(X)\equiv\sum_{\sigma \in S_N}\prod_{i=1}^N X_{i,\sigma(i)}
\end{align}
is the permanent of a matrix $X$, and $S_N$ represents a permutation group. 
The matrix $U_{S,T}$ is obtained from the unitary matrix $U$ by repeating $t_j$ copies of the $j$th column of $U$ to construct a matrix $U_T$ and then by repeating $s_j$ copies of the $j$th row of $U_T$.
In general, calculating the probability of an outcome is hard because computing permanent is \#P-complete \cite{valiant1979complexity}.
On the other hand, if a system has a number of multiphoton events at each mode (collision), one can expect that the computation of the corresponding permanent becomes easier because the relevant matrix has a simpler structure.
Based on the difficulty of calculating permanent when multiphoton events are suppressed by assuming $M\geq  N^6$, it has been proven that the existence of a classical computer that is able to efficiently simulate the boson sampling leads to collapse of the polynomial hierarchy (PH) under some conjectures \cite{aaronson2011computational}.

While the hardness proof of classical simulation of boson sampling assumes an ideal bosonic quantum device, there are various inevitable imperfections in boson sampling experiments \cite{wang2019boson}.
Photon-loss is one of the most critical imperfections in quantum optics experiments, which can be described by the transformation of a mode operator as shown in Fig. \ref{fig:circuit} (c),
\begin{align}
\hat{a}\to \sqrt{\mu}\hat{a}+\sqrt{1-\mu}\hat{e},
\end{align}
where $\hat{e}$ denotes the mode operator of the environment, and $\mu$ denotes the transmissivity.
We assume the environment's quantum state to be in a vacuum state, which is a reasonable assumption in optical frequency.
The photon-loss model can describe imperfect preparation of single photons and inefficiency of single-photon detectors as well as photon-loss in the circuits.
In addition, it is natural to assume that the photon-loss rate is the same on each mode in practice. Note that our MPO algorithm is also applicable to non-uniform loss \cite{brod2020classical} although it requires more computational time (See Sec. \ref{appendix:simul}).

Especially in the uniform loss case, one can easily verify that photon-loss channels commute with arbitrary beam-splitter circuits as shown in Fig. \ref{fig:circuit} (d).
Thus, denoting $\mu_p$, $\mu_u$, and $\mu_m$ as the transmissivity for each photon-loss on preparation, unitary circuits, and measurement, the total transmissivity is given by their product $\mu=\mu_p\mu_u\mu_m$. 
As a result, the uniform photon loss can be captured by combining all the photon loss into photon-loss only on the preparation step such that each single-photon state is replaced by
\begin{align}\label{ini_state}
|1\rangle\langle1|\to \hat{\sigma}=(1-\mu)|0\rangle\langle 0|+\mu|1\rangle\langle1|.
\end{align}
Thus, we now assume that the rest of the process, such as beam splitters and measurement, is perfect as shown in Fig. \ref{fig:circuit} (b).

It is worth emphasizing that the use of one-dimensional (1D) architecture in this work is only for simulation purposes. That is, while we use 1D architecture, we choose a sufficiently large circuit depth $D\approx M$ so that it generates a passive unitary circuit $\hat{U}$ that mixes mode operators via a \textit{global} $M\times M$ Haar-random unitary matrix. Hence, our results apply to any architecture [including two-dimensional (2D) architectures, e.g., used in Ref.\ \cite{wang2019boson}, and the ones with more complex connectivity] that aims to realize a global Haar-random unitary matrix. In particular, our result is independent of the choice of architecture because we are only interested in how many photons go into a Haar-random passive circuit (i.e., $N$) and how many photons are detected by the photon-number detectors (i.e., $N_{\textrm{out}}$). Lastly, we remark that while we arbitrarily put $N$ input photons to the first $N$ modes, our results are independent of this arbitrary choice because of the Haar-random nature of the unitary matrix $U$.

\section{Method}\label{sec:method}
\subsection{MPS simulation} \label{sec:mps}
In this section, we introduce an MPS method to simulate boson sampling \cite{vidal2003efficient}.
MPS is a useful tool to represent a quantum state of a many-body system.
The canonical form of an MPS representation \cite{schollwock2011density} is written as
\begin{align}
|\psi\rangle&=\sum_{i_1,\cdots,i_M=0}^{d-1}c_{i_1\cdots i_M}|i_1,\cdots, i_M\rangle\nonumber  \\ 
&=\sum_{i_1,\cdots,i_M=0}^{d-1}\sum_{\alpha_0,\cdots,\alpha_{M}=0}^{\chi-1}\Gamma_{\alpha_0\alpha_1}^{[1]i_1}\lambda_{\alpha_1}^{[1]}\Gamma_{\alpha_1\alpha_2}^{[2]i_2}\lambda_{\alpha_2}^{[2]} \nonumber \\ 
&~~~~~~~~~~~~~\times \cdots \lambda_{\alpha_{M-1}}^{[M-1]}\Gamma_{\alpha_{M-1}\alpha_{M}}^{[M]i_M}|i_1,\cdots, i_M\rangle,
\end{align}
where $d$ is the dimension of a local Hilbert space and $\chi$ is the bond dimension.
Here, the vectors $\lambda_{\alpha_k}^{[k]}$ represent the singular values in a spectral decomposition for bipartitions, $|\psi\rangle=\sum_{\alpha_k=0}^{\chi-1}\lambda_{\alpha_k}^{[k]}|\psi^{[1,\cdots, k]}_{\alpha_k}\rangle|\psi^{[(k+1),\cdots, M]}_{\alpha_k}\rangle$.
Also, bond dimension can be understood as the maximum Schmidt rank over all bipartitions \cite{vidal2003efficient}.
Thus, we need a large number of bond dimension when a quantum state is more entangled.
We provide details of the standard MPS representation and how to update the MPS after applying two-site gates in Appendix \ref{appendix:mpsmpo}.

While an arbitrary quantum state can be described by an MPS, the time and memory cost for an MPS depends on its bond dimension $\chi$.
More precisely, the memory cost of an MPS is $O[\chi^2dM+\chi(M-1)]$ for tensors $\Gamma$ and $\lambda$.
More importantly, when we apply a unitary operation on a state, the standard update of an MPS requires matrix multiplications and a singular value decomposition, which takes computational time $O(d^4\chi^3)$ and $O(d^3\chi^3)$, respectively.
Thus, by restricting a bond dimension and approximating a given quantum state by choosing the largest $\chi$ singular values for each partition, one can reduce the computational complexity (see Appendix \ref{appendix:mpsmpo} for details).

Since boson sampling circuits are composed of passive transformations, the total system has global U(1) symmetry (photon number preserving), which can be exploited to improve the MPS simulation more efficiently \cite{singh2011tensor, guo2019matrix, huang2019simulating}.
The basic idea is that when the system has U(1) symmetry, MPS tensors can be decomposed into blocks having different photon numbers.
Then a matrix multiplication and a singular value decomposition can be performed for each block with different photon numbers.
Thus, the matrix size for a singular value decomposition is reduced.
We provide the details of how U(1) symmetry reduces the computation time in Appendix \ref{appendix:simul}.

\subsection{MPO simulation}\label{sec:mpo}
An MPS representation can be generalized to describe mixed states \cite{verstraete2006matrix, zwolak2004mixed}.
Basically, we exploit a similar representation to MPS by vectorization of a given density matrix $\hat{\rho}$ such that
\begin{align}
&\hat{\rho}=\sum_{i_1,i_1',\cdots, i_M, i_M'=0}^{d-1}\rho_{i_1,i_1',\cdots, i_M,i'_M}|i_1,\cdots, i_M\rangle\langle i_1',\cdots, i_M'| \nonumber \\
&\to |\hat{\rho}\rangle\rangle
=\sum_{i_1,\bar{i}_1',\cdots,i_M,\bar{i}_M'=0}^{d-1}\sum_{\alpha_0,\cdots,\alpha_{M}=0}^{\chi-1}\Gamma_{\alpha_0\alpha_1}^{[1]i_1\bar{i}_1'}\lambda_{\alpha_1}^{[1]}\Gamma_{\alpha_1\alpha_2}^{[2]i_2\bar{i}_2}\lambda_{\alpha_2}^{[2]} \nonumber \\
&~~~~~~~~~~~~~~~\times \cdots \lambda_{\alpha_{M-1}}^{[M-1]}\Gamma_{\alpha_{M-1}\alpha_M}^{[M]i_M\bar{i}_M'}|i_1,\bar{i}_1',\cdots, i_M, \bar{i}_M'\rangle\rangle.
\end{align}
Here, we have vectorized $|i_j\rangle\langle i_j'|$ to $|i_j, i_j'\rangle\rangle$ for each site.
Note that after the vectorization, an effective local dimension increases from $d$ to $d^2$, which increases the time and memory cost of simulation.
We provide more details of the standard MPO method in Appendix \ref{appendix:mpsmpo}.
Similarly to MPS simulation, the bond dimension $\chi$ determines the computation cost.
The memory requirement is $O[\chi^2d^2M+\chi(M-1)]$ for the tensors $\Gamma$ and $\lambda$, where the local dimension is changed from $d$ to $d^2$.
The time cost for a unitary update is $O(d^8\chi^3)$.
Again, we can employ U(1) symmetry for MPO simulation to reduce the computational cost \cite{guo2019matrix} (see Appendix \ref{appendix:simul} for details).

\subsection{MPS / MPO approximability}\label{sec:mpoee}
As shown in the previous sections, dominant computational time is spent for a singular value decomposition, so that the computational time cost of MPS and MPO simulation is determined by their bond dimension $\chi$. More precisely, the computational time cost is written as $T=O(MDd^4\chi^3)$ for an MPS simulation and $T=O(MDd^8\chi^3)$ for an MPO simulation.
In this section, we introduce a way to determine how the bond dimension scales for a given problem.
Let us focus on an MPS first. 
In general, for the exact description of an arbitrary pure quantum state, $\chi=d^{\left\lfloor{M/2}\right\rfloor}$ number of bond dimension is required, which necessitates an exponential time cost.
In order to avoid exponential computational cost as a system size increases, we fix the bond dimension $\chi$ and approximate a given quantum state by keeping the largest $\chi$  singular values only after the update for unitary operations and discarding the smallest singular values.
When a quantum state's entanglement is limited, the required bond dimension does not increase exponentially \cite{vidal2003efficient}.
More precisely, whether an exponential number of bond dimension $\chi$ is necessary to approximate a given quantum state is determined by MPS and MPO EE, which is introduced as follows.

First of all, an MPS can efficiently approximate a quantum state if the entanglement of the quantum state is not large enough \cite{vidal2003efficient, verstraete2006matrix, schuch2008entropy}.
Formally, if for a family of quantum states of interest $\{|\psi_N\rangle\}$ there exist $c,c'>0$ and $0\leq \alpha <1$ such that $S_\alpha(\hat{\rho}^k_N)\leq c\log_2 N+c'$ for all reduced density matrices $\hat{\rho}^k_N=\text{Tr}_{[1,\cdots, k]}[|\psi_N\rangle\langle\psi_N|]$, then it can be efficiently approximated by an MPS in the sense that the trace distance between an ideal state and an approximate states by an MPS can be made arbitrarily small using $\chi=\text{poly}(N)$ \cite{verstraete2006matrix, schuch2008entropy}.
Here, $S_\alpha(\hat{\rho})$ is the R\'{e}nyi entropy of a density matrix $\hat{\rho}$,
\begin{align}
S_\alpha(\hat{\rho})\equiv\frac{\log_2(\text{Tr}\hat{\rho}^\alpha)}{1-\alpha}, ~~~~ 0\leq \alpha<\infty,
\end{align}
and $\lim_{\alpha\to1}S_\alpha(\hat{\rho})\equiv S_1(\hat{\rho})= -\text{Tr}[\hat{\rho}\log_2 \hat{\rho}]$, is von Neumann entropy.
Note that trace distance is an upper-bound of total variance distance,
\begin{align}
    \frac{1}{2}\sum_x|P(x)-P_a(x)|\leq \frac{1}{2}\|\hat{\rho}-\hat{\rho}_a\|_1,
\end{align}
where $\hat{\rho}$ and $P(x)$ ($\hat{\rho}_a$ and $P_a(x)$) represent an ideal (approximate) density matrix and probability distribution of an outcome $x$ after measurement, respectively.
Thus, when the R\'{e}nyi entropy satisfies the above condition, an MPS with $\chi=\text{poly}(\chi)$ allows an efficient description of the state and sampling (See Appendix \ref{appendix:sample}).
On the contrary, if $S_1(\hat{\rho}^k_N)$ linearly increases or there exists $\alpha>1$ and $\kappa>0$ such that $S_\alpha(\hat{\rho}^k_N)$ increases as $S_\alpha(\hat{\rho}^k_N)\propto cN^\kappa$, an MPS cannot efficiently describe the quantum state, i.e., we need an exponential number of bond dimension $\chi=O(\text{exp}(N^\kappa))$ \cite{schuch2008entropy} to approximate the quantum states.

Therefore, the computational cost of an MPS simulation can be quantified by using MPS EE.
Based on the relation between MPS EE and MPS approximablity, in this work, we investigate the maximum MPS EE over all possible bipartitions,
\begin{align}
S_\alpha^\text{max}(|\psi\rangle)&\equiv \max_{1\leq k \leq M-1}S_\alpha (\hat{\rho}^k=\text{Tr}_{[1,\cdots, k]}[|\psi\rangle\langle\psi|]) \nonumber \\ 
&=\frac{\log_2 \left[\sum_{\beta=0}^{\chi-1}(\lambda_\beta^{[k]})^{2\alpha}\right]}{1-\alpha},
\end{align}
and its behavior as a system size increases.

One can find the same relation for MPO approximability with a minor modification.
The difference of MPO approximation from MPS approximation is that singular value vectors are not necessarily normalized even if the bond dimension $\chi$ is large enough: $\sum_{\alpha_k=0}^{\chi-1} (\lambda_{\alpha_k}^{[k]})^2\neq 1$.
Thus, we first normalize singular value vectors when we initialize an MPO for an input state.
Since the rest of the quantum circuits are unitary operations, the singular value vectors are normalized even after updating for quantum circuits as long as a bond dimension is chosen large enough (see Appendix \ref{appendix:mpsmpo}).
Therefore, the same relation between MPO EE and MPO approximability holds for MPO simulation by defining the maximum MPO EE as
\begin{align}
S_\alpha^\text{max}(|\hat{\rho}\rangle\rangle)\equiv \max_{1\leq k \leq M-1}S_\alpha(\text{Tr}_{[1,\cdots, k]}[|\hat{\rho}\rangle\rangle\langle\langle\hat{\rho}|]).
\end{align}
Note that for pure states, MPO EE is equal to twice MPS EE because a vectorized pure state simply represents two equivalent pure states, increasing the local dimension from $d$ to $d^2$.
It is worth emphasizing that because of vectorization, approximation accuracy is defined as the vector 2-norm between vectorized ideal and approximate states, which is equal to matrix 2-norm between ideal and approximate density matrices rather than trace distance \cite{jarkovsky2020efficient}.
Because of the relation between matrix 1-norm and 2-norm, $K\|A\|_2\geq \|A\|_1$,
where $K$ is the dimension of the matrix $A$, MPO EE may decrease even if a larger bond dimension is required to bound matrix 1-norm between an ideal quantum state and an approximated state of an MPO.
On the other hand, it is guaranteed that an MPO is inefficient if R\'{e}nyi (von Neumann) entropy of $\alpha>1$ algebraically (linearly) increases.

\begin{figure*}[t!]
\includegraphics[width=450px]{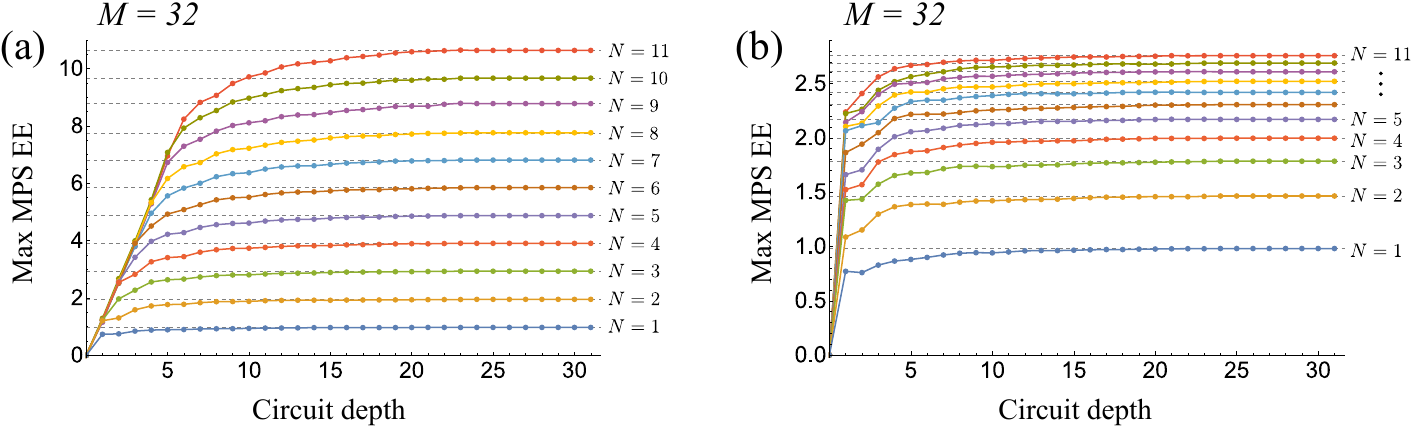}
\caption{Maximum MPS EE $(\alpha=1)$ for different number of photons from $N=1$ to $11$ from the bottom to the top with different circuit depths. The number of modes is $M=32$. (a) An input state is $N$ single photons in different modes. (b) An input state is $N$ photons in the first mode. Note that the bipartition that leads to the maximum MPS EE at the last step is always $[1\cdots 16]:[17\cdots 32]$ and that the circuit depth is defined as in Appendix \ref{appendix:Haar-random}.}
\label{fig:MPS2}
\end{figure*}

\section{Results}\label{sec:results}
\subsection{Lossless Boson Sampling}\label{sec:lossless}
First, we numerically simulate lossless boson sampling with an input state $|\psi_\text{in}\rangle=|1\rangle^{N}|0\rangle^{M-N}$ using MPS.
Since we use $N$ single photons, the dimension of local Hilbert spaces is $d=N+1$.
We first initialize an MPS, update the MPS according to a unitary circuit composed of Haar-random beam splitters, and finally obtain the output state before measurement and calculate the maximum MPS EE over bipartitions.
We repeat the procedure with different circuits to obtain the average of the maximum MPS EE.

In order to minimize the depth of a Haar-random circuit, we have used the fact that any $M$-mode passive unitary transformation can be decomposed into $M(M-1)$ number of beam splitters \cite{reck1994experimental, clements2016optimal} and that a Haar-random circuit can be obtained by sampling the transmissivities of the beam splitters in a structured manner \cite{russell2017direct} with a depth $D\approx M$ (See Appendix B for details).

Using an MPS simulation procedure as introduced in Sec. \ref{sec:mps}, we first simulate the boson sampling with a fixed number of modes $M=32$ and different input photon numbers from $N=1$ to $N=11$.
We take the average of the maximum MPS EE over 200 different circuit configurations for $1\leq N \leq 9$ and 10 different circuits for $N=10, 11$.
One can show that the optimal bond dimension needed to implement an MPS simulation without truncation error is given by $\chi=2^{N}$ and that typical random circuits make MPS EE close to $N$ for large $M$ (See Appendix \ref{appendix:entropy}).
Furthermore, when $N$ and $M$ are sufficiently large, we show that MPS EE increases linearly in $N$ in Appendix \ref{appendix:entropy}.
The underlying principle is that for a given bipartition $[1 \cdots l]:[(l+1)\cdots M]$, since each single photon occupies either partition after beam splitter arrays, we need the bond dimension $\chi=2^N$ to describe an output state without any truncation error.

Figure \ref{fig:MPS2} (a) indeed shows that the maximum MPS EE with $\alpha=1$ linearly increases as the number of photons in the system.
The linearly increasing maximum MPS EE implies that an exponential number of bond dimension is necessary to simulate the boson sampling within a desired accuracy as the number of single photons increases.
Notice that in this simulation, the number of input photons $N$ is not much smaller than the number of modes $M$, whereas in the original proposal of boson sampling \cite{aaronson2011computational}, the number of modes is assumed to be much larger than the number of photons, namely $M\geq N^6$, to prove the hardness of boson sampling.
Nevertheless, we obtain a constant difference of maximum MPS EE between a successive number of input photons.
Therefore, it suggests that the MPS simulation is inefficient even if the number of modes is not large enough compared to the number of input photons.
We note that circuit depth in Fig. \ref{fig:MPS2} is defined slightly differently than a standard way (See Appendix \ref{appendix:Haar-random}).

\begin{figure}[b]
\includegraphics[width=200px]{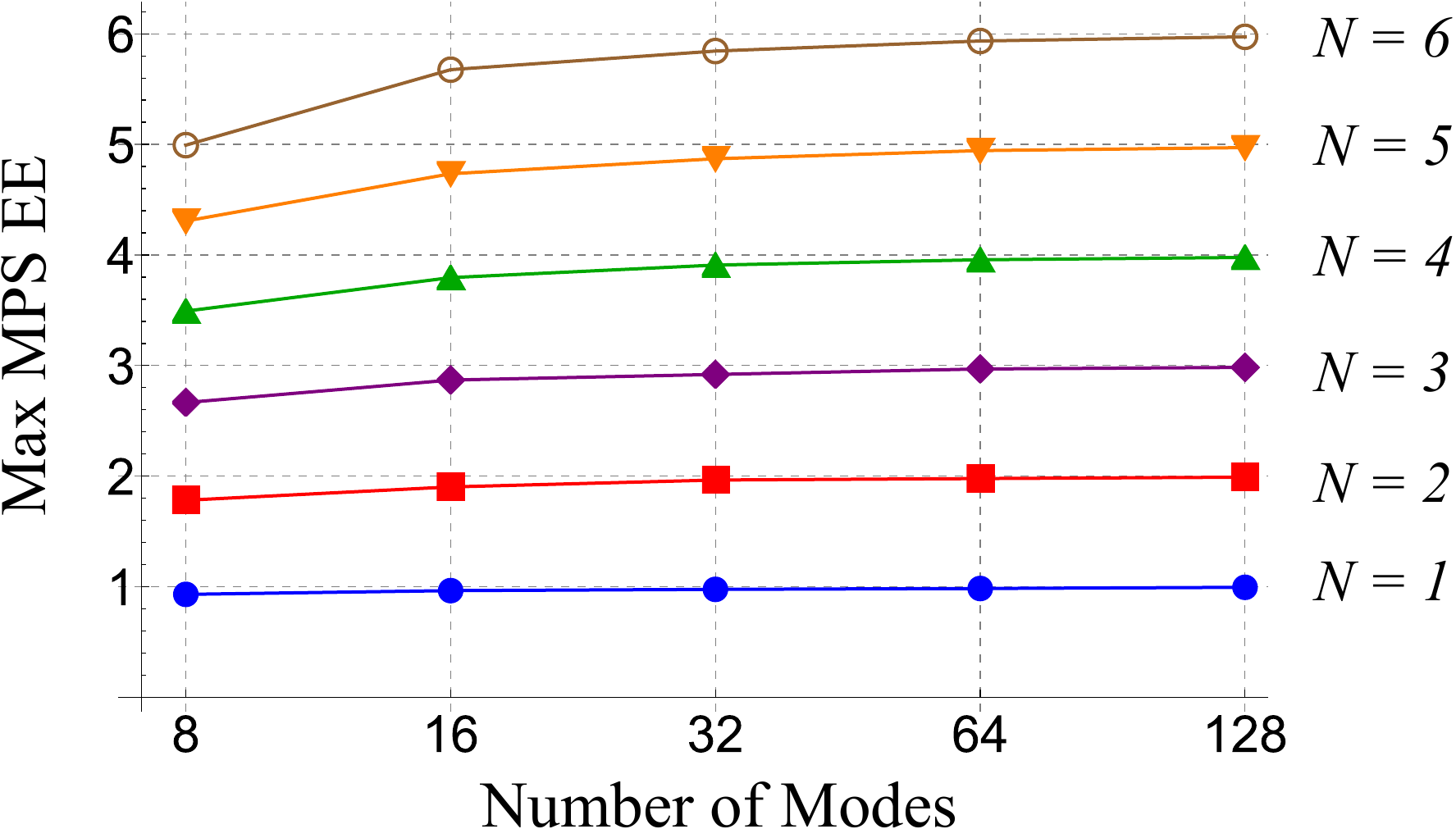}
\caption{Maximum MPS EE $(\alpha=1)$ for different number of modes $M$ and photons $N$. We averaged 50 different circuits to obtain each point. Note that the bipartition that leads to the maximum MPS EE is always the center.}
\label{fig:MPS}
\end{figure}

We compare the MPS simulation of the standard boson sampling \cite{aaronson2011computational} with the one with a different type of an input state $|\psi_\text{in}\rangle=|N\rangle|0\rangle^{M-1}$ (Note that more general input states are analyzed in Ref. \cite{brod2020classical} and Appendix \ref{appendix:entropy}.).
It can be easily shown that the computation of the probability of an outcome in this case is not difficult because the corresponding permanent is constructed by repeating $N$ times of the same column \cite{aaronson2011computational}.
In the simulation, we take the average of the maximum MPS EE over 200 different circuit configurations for $1\leq N \leq 11$.
Again, we have shown that the bond dimension needed to implement MPS simulation without truncation error is $\chi=N+1$, which already implies that an efficient simulation is possible (See Appendix \ref{appendix:entropy}).
In this case, in contrast to the previous case, the photons do not behave independently, so that an exponential number of bond dimension is not required.
Indeed, we show that MPS EE increases logarithmically as the number of photons grows in the same mode in Appendix \ref{appendix:entropy}.
Figure \ref{fig:MPS2} (b) shows a different behavior of the maximum MPS EE from the previous standard boson sampling.
As expected, in contrast to the previous case, the maximum MPS EE does not increase linearly and the gap of the maximum MPS EE between a successive number of photons decreases as the input photon number increases.

Finally, we analyze the maximum MPS EE for a fixed number of input photons and different number of modes, which is shown in Fig. \ref{fig:MPS}.
We average over 50 different circuit configurations.
Interestingly, as we increase the number of modes in the circuit for a fixed input photon number $N$, the maximum MPS EE converges to $N$.
Thus, if the number of modes is large enough, increasing the number of modes no longer makes the MPS simulation hard, which is consistent with the Clifford-Clifford algorithm where the time cost for simulation is $T=O[N2^N+\text{poly}(M,N)]$; the complexity in terms of the number of modes $M$ is polynomial \cite{clifford2018classical}.
In addition, recently it was shown that when the number of modes is proportional to the number of photons, $M\propto N$, the classical simulation can be faster than when the number of modes is much larger than the number of photons \cite{clifford2020faster}.
Especially when $M=N$, the computational cost of boson sampling is $T=O(N \rho^N+N^3)$ with $\rho=27/16 \approx 1.69$.
Our MPS simulation also shows a similar tendency that the difference of the maximum MPS EE of two successive input photon numbers gets smaller when the number of modes is small.

\subsection{Lossy Boson Sampling}\label{sec:lossy}
In this section, we analyze the effect of photon-loss in boson sampling circuits by investigating maximum MPO EE.
As previously mentioned in Sec. \ref{sec:bs}, we introduce photon-loss by using imperfect single-photon sources $\hat{\sigma}$:
\begin{align}
\hat{\rho}=\hat{\sigma}^{N}\otimes |0\rangle\langle 0|^{M-N}. 
\end{align}
We denote $N_\text{out}=\mu N$ as a total mean photon number after loss channel, where the transmission rate $\mu=\mu(N)$ is a function of the number of input photons $N$.
In other words, we study the relation between hardness of the simulation and a loss rate depending on the input photon number.
Before we present our main results, we show why classically simulating lossy boson sampling is nontrivial.

\subsubsection{Complexity of Lossy Boson Sampling}
First, we note that exact simulation of lossy boson sampling is hard unless the PH collapses: suppose that we have a classical simulator that can efficiently and exactly simulate lossy boson sampling. Then, with this simulator, one can exactly simulate a lossless boson sampling as well by post-selecting the case where no photon is lost. 
However, since an exact boson sampling with a post-selection allows a universal quantum computation, the existence of an efficient and exact lossy boson sampler implies that $\text{PH}\subseteq \text{P}^\text{PP}=\text{P}^\text{post-BQP}=\text{P}^\text{post-BPP}$ \cite{aaronson2005quantum}, which contradicts to the fact that $\text{P}^\text{post-BPP}$ is in the PH \cite{han1997threshold} assuming that the PH is infinite.
Therefore, we focus on an approximate simulation of lossy boson sampling.

We emphasize that using a classical algorithm ideal boson sampling in a trivial way does not significantly reduce the complexity and that a potentially efficient classical lossy boson sampler should systematically employ the fact that loss makes the entanglement grow slower.
We now explicitly present an algorithm that employs a classical boson sampler to simulate a lossy boson sampling in a naive way.
Assume that we have a classical boson sampler that takes an exponential computational time $c^n (c>1)$ to simulate the ideal boson sampling with $n$ single photons (e.g., the Clifford-Clifford algorithm \cite{clifford2018classical}).
Since a single-photon state after a loss channel is a mixture of vacuum $|0\rangle$ and a single-photon state $|1\rangle$ with a probability $1-\mu$ and $\mu$, respectively, one may sample a pure input photon configuration from a binomial distribution for the first $N$ input modes with a transmission rate $\mu$ and execute a classical boson sampler using the sampled input state.
If the procedure is iterated for a number of samples, the average time cost can be given by
\begin{align}
    T&=\sum_{n=0}^N \binom{N}{n}\mu^n (1-\mu)^{N-n} c^{n}=\left[1+\mu(c-1)\right]^N. 
\end{align}
Especially when the loss-scaling follows a power-law such that $N_\text{out}=\beta N^{\gamma}~(0< \gamma< 1)$, the time cost in an asymptotic regime is simplified as
\begin{align}
    T=\left[\left(1+\beta \frac{c-1}{N^{1-\gamma}}\right)^{N^{1-\gamma}}\right]^{N^{\gamma}}
    \to e^{(c-1)N_\text{out}}.
\end{align}
Thus, such a simple procedure using binomial sampling and a classical boson sampler requires an exponential time cost because it pursues exact simulation of a lossy boson sampling.

On the other hand, one may choose only dominant binomial coefficients in binomial sampling for approximate sampling.
Since the dominant binomial coefficients are around $N_\text{out}$, the computational cost to run the classical boson sampler is given by $O(c^{N_\text{out}})=O(c^{N^\gamma})$, which is still inefficient for $0<\gamma<1$.
Even though it decreases the complexity, such an approach does not fully exploit the fact that the system is lossy because it still samples a pure state to run an ideal classical boson sampler.
By contrast, a potentially more efficient classical algorithm for lossy boson sampling should properly employ the fact that the output state from which we sample is a mixture of pure quantum states.
The mixedness makes the output state less entangled than an output state in lossless boson sampling, which is the key to reduce the complexity of lossy boson sampling.
For this reason, an MPO simulation that we propose has a major advantage for lossy boson sampling since it systematically exploits the fact that loss in the system makes the entanglement grow slower.

Indeed, there have been many proposals of an efficient approximate classical algorithm for lossy boson sampling using the mixedness of the output state.
Particularly, an efficient approximate classical simulation algorithm for a scaling $N_\text{out}\propto\sqrt{N}$ has been proposed when a loss rate is large or in an asymptotic regime in Refs. \cite{oszmaniec2018classical, garcia2019simulating}.
The proposed simulation is based on finding the closest thermal state \cite{garcia2019simulating} or the closest particle-separable state \cite{oszmaniec2018classical}, which can be used for an efficient simulation.
However, because the simulations rely on a particular state determined by given parameters, providing more time for the simulation does not improve its accuracy.
Unlike the previous studies, one can efficiently control our MPO simulation's accuracy by adjusting the bond dimension in the simulation.
Also, our MPO simulation focuses on the behavior of classical simulation of lossy boson sampling a non-asymptotic regime where a loss rate is not very large so that thermal states fail to approximate the output state properly.

We remark that another way to approximately simulate lossy boson sampling is to discard the probability of outcomes corresponding to a large degree of multiphoton interference, which is highly suppressed when the system is lossy \cite{renema2018classical}.
Thus, choosing a threshold of the degree of multiphoton interference allows one to control an approximation error.
On the other hand, our MPO simulation controls a simulation error by keeping dominant singular values and discarding small singular values.

\begin{figure*}[t!]
\includegraphics[width=500px]{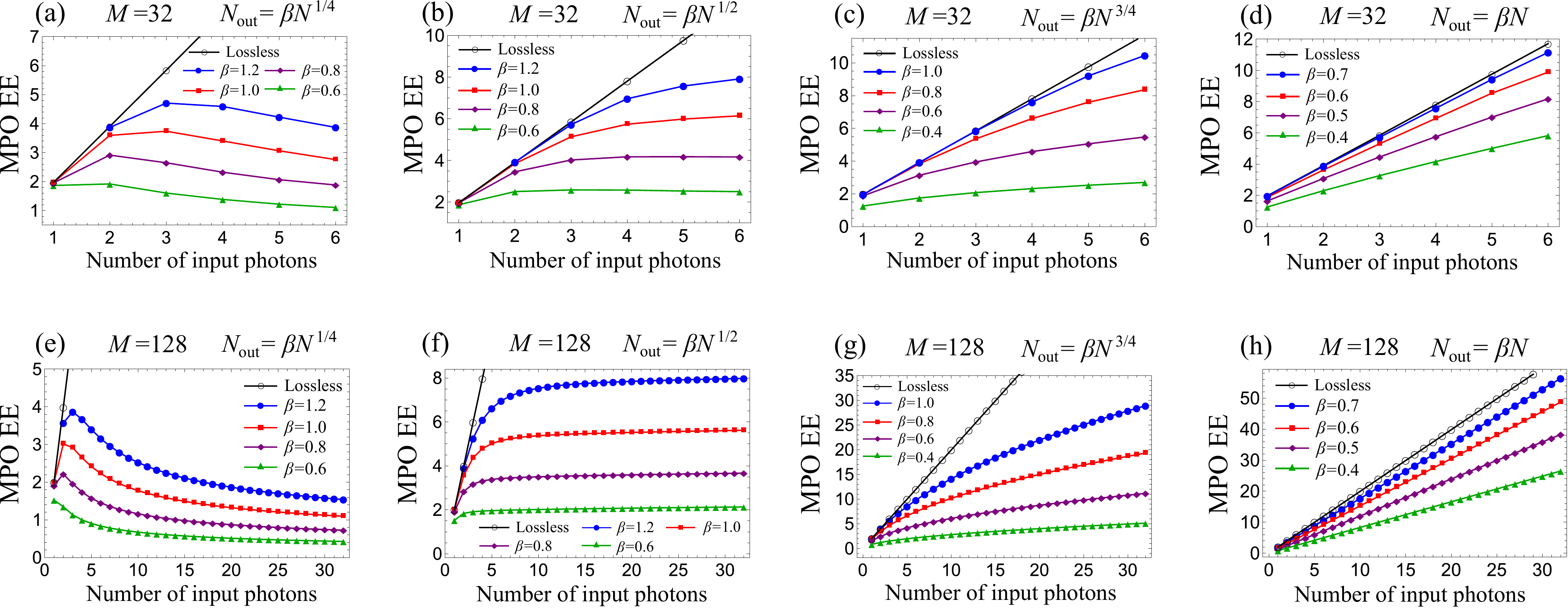}
\caption{Maximum MPO EE obtained by simulation for $M=32$ and different input photons numbers $N=1\sim 6$ and different loss scales of (a) $N_\text{out}=\beta N^{1/4}$, (b) $N_\text{out}=\beta\sqrt{N}$, (c) $N_\text{out}=\beta N^{3/4}$, and (d) $N_\text{out}=\beta N$. Empty circles represent the maximum MPO EE for lossless case. (d) $N_\text{out}=\beta N$. Maximum MPO EE directly computed for $M=128$ and different input photons numbers $N=1\sim 32$ and different loss scales of (e) $N_\text{out}=\beta N^{1/4}$, (f) $N_\text{out}=\beta\sqrt{N}$, (g) $N_\text{out}=\beta N^{1/4}$, and (h) $N_\text{out}=\beta N$. Empty circles represent the maximum MPO EE for lossless case. Note that the difference of MPO EEs between the upper panel and the lower panel for the same parameters is present because we simulated with U(1) symmetry to obtain MPO EE for (a)-(d) and computed MPO EE without the symmetry for (e)-(h).
}
\label{fig:mpo}
\end{figure*}

\subsubsection{MPO EE for various loss scalings}
We first simulate the case where a loss-rate is constant in the number of input photons, i.e., $N_\text{out}=\mu N (\gamma=1)$ with a constant $0<\mu<1$.
Figure \ref{fig:mpo} (d) shows that in this case, an maximum MPO EE ($\alpha=1$) linearly increases as $N$.
It indicates that boson sampling for a constant loss rate cannot be efficiently simulated using MPO because a bond dimension for an accurate approximation is required to increase exponentially as an input photon number increases.
To the best of our knowledge, a constant loss case has not been investigated yet, and our numerical result provides evidence that hardness of boson sampling might persist in this regime.
Here, the average of the maximum MPO EE is taken over 100 different circuits for $1\leq N \leq 5$ and 10 different circuits for $N=6$.
The maximum bond dimension we used is $\chi=4000$.
Note that even though we fix the number of modes to be $M=32$ throughout the simulation, we have checked for MPO simulation that increasing the number of modes further does not change the MPO EE much, similarly to Fig. \ref{fig:MPS}.

A constant loss-scaling is particularly important from an experimental perspective although it is a rather optimistic scaling.
First, when one increases the number of single photons with fixing number of modes, the loss rate for the whole optical circuits can be assumed to be constant because we assume a uniform loss on each mode.
However, experimentally various factors will degrade the performance of boson sampling such as diminishing of distinguishability of single photons and a coincidence detection rate.
In addition, when the number of modes increases as the number of photons as the original proposal \cite{aaronson2011computational} and the depth of the circuit to implement a Haar-random unitary matrix accordingly, it becomes more difficult to maintain the same loss rate.
Nevertheless, our numerical results indicate that if one can manipulate the loss rate for the entire circuit to be constant with increasing the number of photons, classical simulations for the loss-scaling might be inefficient.
We emphasize that more rigorous complexity-theoretical proof is required.

Since the above scaling is somewhat optimistic in practice, we analyze a scaling where a loss rate increases as the number of single photons ($\gamma<1$).
If the depth of a circuit increases as the number of input photons following the original proposal \cite{aaronson2011computational}, the loss rate of the entire circuit will increase accordingly.
An interesting scaling is $N_\text{out}=\mu N=\beta\sqrt{N}$, where an efficient simulation with a constant error has been proposed \cite{oszmaniec2018classical, garcia2019simulating}.
Remarkably in this scaling, one can observe that for a small coefficient $\beta$, the maximum MPO EE ($\alpha=1$) saturates or even decreases when $N_\text{out}$ increases, which is shown in Fig. \ref{fig:mpo} (b).
One can observe that the behavior is clearly different from lossless cases or $\gamma=1$ cases.
The simulation result suggests that the computational cost of an MPO simulation for lossy boson sampling does not increase as fast as lossless boson sampling.
The tendency is more apparent when $\gamma<1/2$.
For example, when $\gamma=1/4$ as shown in Fig. \ref{fig:mpo} (a), the maximum MPO EE decreases for a broad range of $\beta$.
On the other hand, when $\gamma=3/4$, although it is slower than linear, the MPO EE increases fast enough to be hard to simulate using polynomial number of a bond dimension as shown in Fig. \ref{fig:mpo} (c).
We emphasize again that even if MPO EE decreases, it does not imply that the computational cost, or the bond dimension, to achieve a desired accuracy for total variance distance reduces because MPO EE is relevant to the vector 2-norm of ideal and approximate vectorized states, while total variance distance may have an extra multiplicative prefactor to the vector 2-norm increasing with the Hilbert space's dimension \cite{jarkovsky2020efficient}.
Nevertheless, the behavior of MPO EE for different loss scalings shows that lossy boson sampling leads to a different tendency of MPO EE from lossless boson sampling.

To analyze MPO EEs of large system size circuits, we use a different approach.
Instead of simulating the circuit using MPO, we directly calculate MPO EE.
Note that in this case, U(1) symmetry is not applied; thus, the values of MPO EE are different from those from simulations.
Figure~\ref{fig:mpo} (e)-(h) present MPO EEs for different loss-scalings and photons numbers for a bipartition $[1\cdots M/2:(M/2+1) \cdots M]$, which gives the maximum over other bipartitions.
MPO EEs are obtained by averaging over $100$ different circuit configurations.
Specifically, we first sample a Haar-random unitary matrix and compute MPO EE, assuming a collision-free case (See Appendix \ref{appendix:entropy} for details).
Notice that collision-free cases give a larger MPO EE than when there are collision events in general.
As expected, when $\gamma<1/2$, the MPO EE decreases as the input photon number increases, which indicates a possibility of an efficient simulation.
In contrast, when $\gamma\geq 1/2$, the MPO EE increases as the input photon number increases, while the increase is slow for $\gamma=1/2$.
Thus, the MPO EE increases extensively for $\gamma>1/2$ so that the entanglement is large enough to simulate efficiently using MPO methods.
Moreover, in a large system size of $N$ and $M$ and for collision-free cases, an asymptotic expression of MPO EEs can be obtained, which is given by (See Appendix \ref{appendix:entropy}),
\begin{align}
    S_\alpha^{M/2}(|\hat{\rho}\rangle\rangle)&= O(N^{1-2(1-\gamma)\alpha}) ~~~~ \text{when}~~~~ \alpha\neq 1, \\ 
    S_1^{M/2}(|\hat{\rho}\rangle\rangle)&= O(N^{2\gamma-1}\log_2{N}).
\end{align}
It shows that when $\gamma<1/2$, MPO EE with $\alpha\to 1$ converges to zero in an asymptotic limit, while there exists $\alpha$ such that MPO EE decreases.
On the other hand, when $\gamma>1/2$, one can find $\alpha>1$ such that MPO EE increases algebraically and conclude that an MPO simulation requires an exponential computational time \cite{schuch2008entropy}.

\begin{figure}[b]
\includegraphics[width=220px]{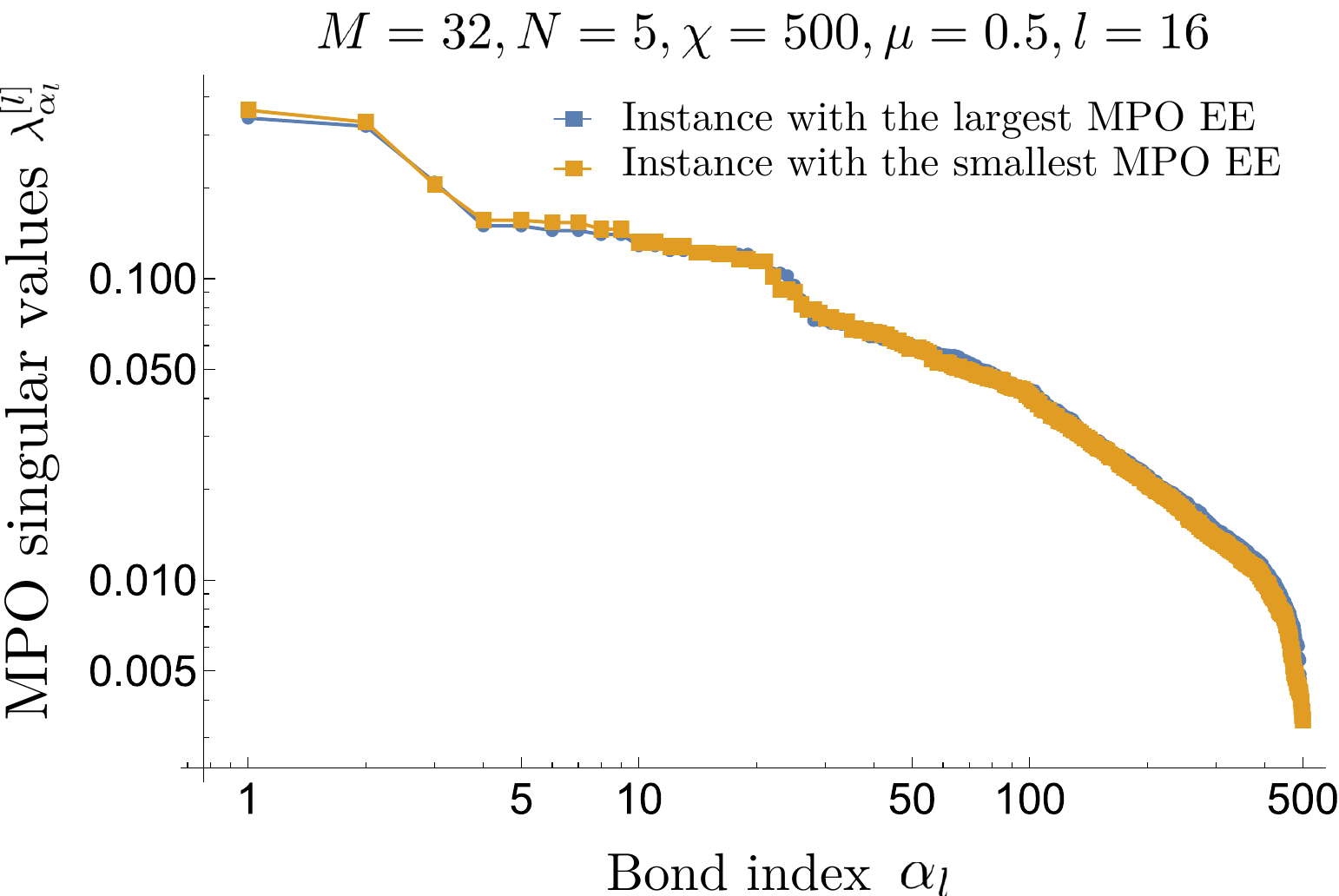}
\caption{Singular value distributions in the descending order for $M=32$ modes, $N=5$ input photons, and bipartition $[1,\cdots, 16]:[17,\cdots, 32]$ with loss rate $\mu=0.5$. The singular value vectors are chosen by the circuits that render the largest and smallest MPO EE among 100 different circuits. Since the singular values decay superpolynomially, the required bond dimension of $\text{poly}(1/\epsilon)$ is enough to achieve an error $\epsilon$.}
\label{fig:SV2}
\end{figure}

\begin{figure*}
\includegraphics[width=480px]{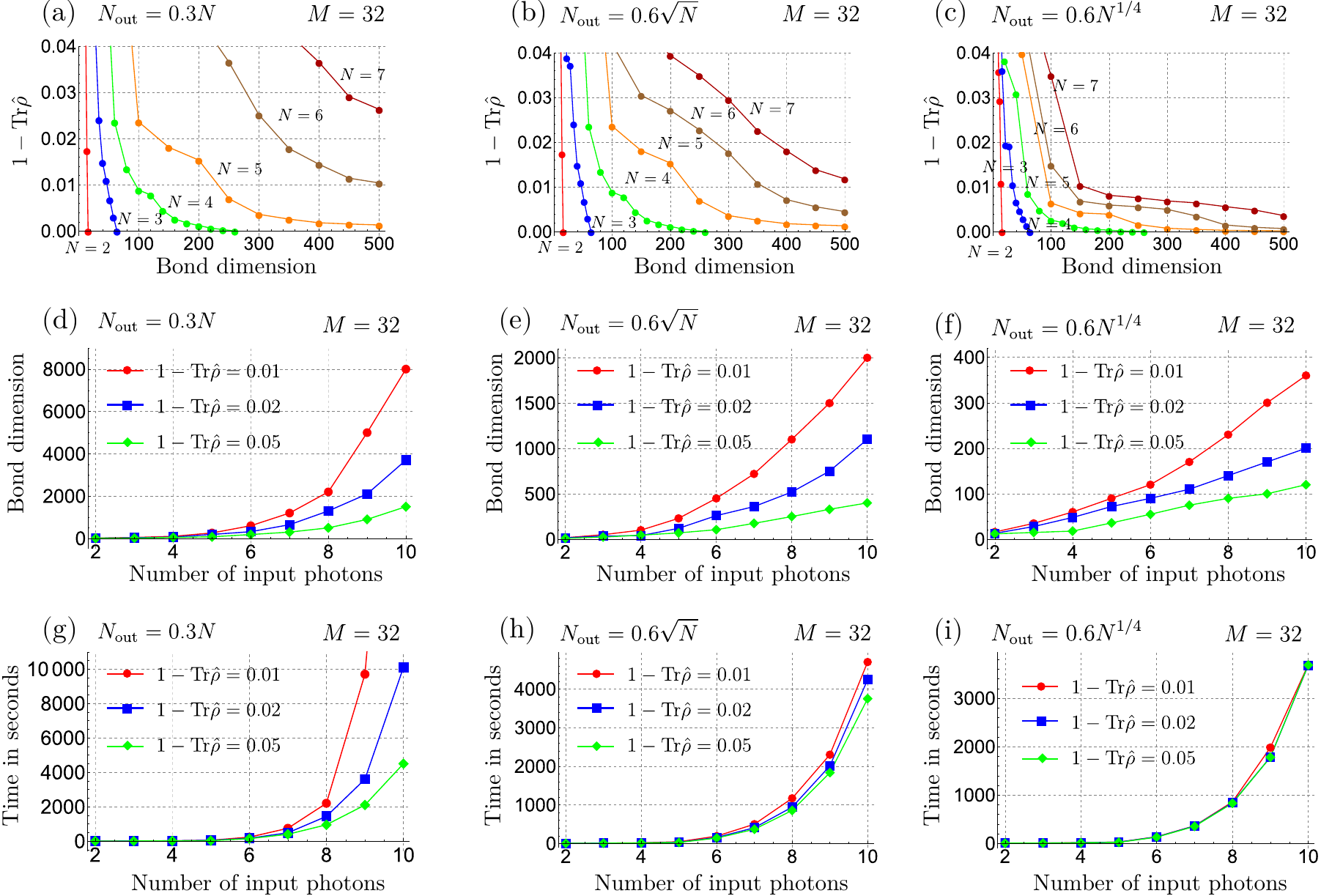}
\caption{MPO simulation errors $1-\text{Tr}\hat{\rho}$ with different bond dimensions for (a) $N_\text{out}=0.3N$, (b) $N_\text{out}=0.6\sqrt{N}$, and (c) $N_\text{out}=0.6N^{1/4}$. The number of modes is $M=32$ and the number of input photons is $N=2,3,4,5,6,7$ (from left to right). 
Bond dimension to achieve errors $1-\text{Tr}\hat{\rho}=0.01, 0.02, 0.05$ for (d) $N_\text{out}=0.3N$, (e) $N_\text{out}=0.6\sqrt{N}$, (f) $N_\text{out}=0.6N^{1/4}$. Running time to attain errors $1-\text{Tr}\hat{\rho}=0.01, 0.02, 0.05$ for (g) $N_\text{out}=0.3N$, (h) $N_\text{out}=0.6\sqrt{N}$, and (i) $N_\text{out}=0.6N^{1/4}$.
The curves are guides for dots. The running time for $N=10$ in (g) is 35000 s.} 
\label{fig:tr}
\end{figure*}

\subsubsection{Relation between simulation accuracy and running time}
We now show that our simulation can effectively improve its accuracy by increasing the bond dimension.
Figure \ref{fig:SV2} presents the distributions of singular values for the case of $M=32$, $N=5$, $\mu=0.5$, $\chi=500$, and the bipartition $[1\cdots 16]:[17\cdots 32]$ as an example.
It shows that the tail of the singular value distribution decreases superpolynomially for the two extreme instances of the largest and smallest MPO EE.
Thus, the bond dimension truncation's impact on the simulation is negligible as long as the bond dimension is chosen such that $\log_2\chi$ is much larger than the MPO EE.
More explicitly, the superpolynomially decaying tail indicates that the required bond dimension $\chi$ and the simulation time cost would increase slower than $\text{poly}(1/\epsilon)$ with $\epsilon$ being the sum of discarded singular values.
We have checked for different parameters and observed the same behavior.
We note that a previously proposed algorithm \cite{garcia2019simulating}, approximating a lossy single-photon state by a thermal state, has an upper bound of total variance distance to be $\beta^2$ for $N_\text{out}=\beta\sqrt{N}$ with an arbitrary $N$, while an MPO simulation's accuracy can be easily controlled.
Especially when $\beta\geq 1$, the former algorithm's total variance distance becomes larger than 1, which indicates that its simulation error might not be bounded properly and shows an MPO simulation's advantage over the algorithm.

More explicitly, we compare how an MPO simulation's accuracy changes as a bond dimension $\chi$ increase for different loss-scalings in Fig. \ref{fig:tr}.
We have used $M=32$ modes with two different loss-scales $N_\text{out}=0.3N$ in Fig. \ref{fig:tr} (a) and $N_\text{out}=0.6\sqrt{N}$ in Fig. \ref{fig:tr} (b).
We have already checked in Fig. \ref{fig:mpo} that MPO EE linearly increases in the former case, whereas it decreases in the latter case as $N$ increases.
In this figure, we measure the error of the simulation as $1-\text{Tr}\hat{\rho}$ instead of total variance distance because total variance distance requires very large computational time for large photon numbers. 
We have checked that $1-\text{Tr}\hat{\rho}$ and total variance distance present a very similar behavior in a small size.
Therefore, we quantify an error here by the amount of lost probabilities from truncation.

First of all, Figures \ref{fig:tr} (a)-(c) show that for a given input photon number $N$, a simulation error can be effectively reduced by increasing a bond dimension.
Moreover, one can observe by comparing between Figs. (a)-(c) that when $N_\text{out}$ grows slowly as $N$, the increment of the required bond dimension becomes smaller, the behavior of which is elaborated in Figs. \ref{fig:tr} (d)-(f).
Also, we present the running time of simulating lossy boson sampling in Figs \ref{fig:tr} (g)-(i).
Figure \ref{fig:tr} (g) clearly shows that when the simulation accuracy is smaller, the running time can be significantly reduced.
Thus, one can more efficiently simulate a lossy boson sampling when $N_\text{out}$ increases slowly and when a target accuracy is smaller.

In Figs. \ref{fig:tr} (h) and (i), the difference of running time for different errors is not significant because the dimension of a matrix for which we perform matrix multiplication and singular value decomposition is small due to U(1) symmetry, so that most of time is spent to employ U(1) symmetry.
We note that the computational overhead to employ U(1) symmetry is $\text{poly}(N)$, which is shown in Appendix \ref{appendix:simul}.
On the other hand, in Fig. \ref{fig:tr} (g), the difference becomes substantial because bond dimension for each charge gets larger, so that matrix multiplication and singular value decomposition are dominant than the overhead for U(1) symmetry.
We note that even if MPO EE decreases for the cases of $N_\text{out}=0.6N^{1/4}$ and $N_\text{out}=0.6\sqrt{N}$ as shown in Figs. \ref{fig:mpo} (a) and (b), a bond dimension and running time to attain a target accuracy can increase, which stems from the fact that MPO EE quantifies a distance between an MPO and an exact state in a vectorized form.
Here, we have used 28 cores of Intel E5-2680v4 2.4GHz to attain the running time in Figs. \ref{fig:tr} (g)-(i).
We note that the size of a matrix that we perform singular value decomposition without using U(1) symmetry is $d^2\chi \times d^2\chi$ with $d=N+1$, which becomes almost intractable quickly as $\chi$ and $N$ increase.

Lastly, we briefly compare our analysis of boson sampling with a related previous work on 1D noisy RCS \cite{noh2020efficient}. Both studies build on an observation that noise tends to reduce non-trivial correlation in quantum systems and use MPOs to more efficiently simulate such noisy systems than the brute force methods. However, while the previous work on RCS is applicable only to 1D architectures, our boson sampling results are not limited to 1D architectures. In our work, the use of 1D architecture is only for the simulation purpose, i.e., for generating a Haar-random boson sampling interferometer (see Appendix \ref{appendix:Haar-random}). Since all boson sampling experiments are set up to realize a Haar-random passive interferometer, our results apply to all such setups regardless of the geometric connectivity of the system. We also remark that unlike the previous work where each gate was assumed to fail with a non-zero gate error rate, we only consider how many photons remain in the system (i.e., $N_{\textrm{out}}$) at the end of the entire process, compared to the input photon number $N$. Note that input photon loss and detection loss rates (analogous to state preparation and measurement error rates) are expected to not depend on the system size. 
Photon loss within the interferometer (analogous to gate error rates) is in principle also taken into account in our model as they will reduce the output photon number $N_{\textrm{out}}$. Unlike input and detection loss rates, however, such loss rate will be enhanced as the system size increases since then larger interferometer is needed to reach Haar randomness and thus more photons will be lost along the way.

\section{Discussion and Conclusion}\label{sec:conclusion}
As experimental scales of boson sampling have been increasing, characterizing the computational cost of a classical simulation of lossy boson sampling becomes more crucial.
Typically, quantum optics experiments suffer from various imperfections, and critical ones in boson sampling are impurity of single-photon sources, photon-loss in the circuit, and inefficiency of photo-detectors.
Due to the aforementioned imperfections, a state-of-the-art boson sampling experiment \cite{wang2019boson} has used $N=20$ input photons but the largest number of photons they detected is only 14, and the sampling rate was not large enough.
The most recent Gaussian boson sampling experiment also suffers from about 70\% of photon loss \cite{zhong2020quantum}.
On the other hand, except for a recently proposed classical algorithm \cite{renema2018classical} employing a fact that photon-loss reduces quantum interference, many classical algorithms proposed for lossy boson sampling are not designed to simulate an intermediate-sized lossy boson sampling where photon loss is not extremely large \cite{oszmaniec2018classical, garcia2019simulating}. 
To overcome this limitation, we have employed MPOs, which enable us to simulate lossy boson sampling with a moderate amount of photon-loss.
An important advantage of the MPO algorithm compared to other classical algorithms is that it can improve its accuracy efficiently.
We have numerically shown that a computational time cost as well as a required bond dimension increases at most polynomially in the simulation error.
We note that, in principle, our MPO scheme can also be used to simulate Gaussian boson sampling by truncating the total photon number properly, which determines the dimension of local Hilbert spaces and the total charge of an MPO representation.
In practice, since MPO for Gaussian boson sampling requires a larger local Hilbert space dimension than single-photon boson sampling, its running time would be larger than the latter.

We have studied the effect of photon-loss on MPO's computational cost of simulating lossy boson sampling as input photon number grows for various loss-scalings using MPO EE.
We first show that if a loss rate can be fixed as the number of photons in boson sampling experiments increases, the computational cost of an MPO simulation exponentially increases.
Since our numerical simulation results rely on a particular simulation method, it does not rule out existence of a more efficient classical simulator that can possibly simulate a constant loss-scaling of boson sampling.
Nevertheless, our results will motivate further rigorous studies for the effect of loss on boson sampling.
We have also demonstrated that an exponential cost is required for MPO simulation for loss-scalings $N_\text{out}\propto N^{\gamma}$ with $\gamma > 1/2$ in an asymptotic limit.
On the other hand, when loss is more severe such that $\gamma\leq 1/2$, the complexity of MPO simulation might not increase exponentially because MPO EE increases at most logarithmically.
Although the same scaling has been studied in Refs. \cite{oszmaniec2018classical, garcia2019simulating}, an important distinctive feature is that our simulation can control the simulation accuracy by increasing its running time.
Therefore, our MPO algorithm can be useful to simulate an intermediate scale of lossy boson sampling with achieving a high accuracy.

We emphasize that a sampling task does not require the full description of an output density matrix as an MPO algorithm does.
Therefore, our MPO algorithm inevitably has a computational overhead than direct sampling algorithms, while it provides more information.
The crucial difference between our MPO algorithm and direct sampling algorithms is that the former takes a lot of time to get the description but sampling from it is very efficient, whereas the latter takes much time cost to obtain each sample.
Also, the full description allows computing output probabilities efficiently, whereas direct sampling algorithms generally do not.

On the other hand, the proof of hardness of boson sampling assumes the number of modes $M$ of a circuit to be much larger than the number of single photons $N$ such that $M\geq N^6$ \cite{aaronson2011computational}.
Although the assumption is expected to be compromised to a less demanding condition $M\geq N^2$ \cite{aaronson2011computational}, it is still far beyond a current technology.
For example, the largest scale of boson sampling experiment so far employed $N=20$ input photons with $M=60$ modes \cite{wang2019boson} and that of Gaussian boson sampling used $N=50$ and $M=100$ \cite{zhong2020quantum}, where $N$ is understood as a number of squeezed states.
Clearly, the number of modes used in the experiment is far smaller than the proposed scale, $M\geq N^2$.
In fact, the fastest known classical algorithm shows that the computational cost to simulate boson sampling can be significantly reduced when the number of modes is linear in the number of photons $N$, although an exponential time cost is still required \cite{clifford2018classical, clifford2020faster}.
Our numerical simulation using an MPS shows a similar tendency to the fastest classical algorithm in the sense that MPS EE increases linearly as an input photon number grows while the increment gets smaller when the number of modes is small.
Considering that the requirement of large number of modes is another critical obstacle to hinder one from demonstrating quantum supremacy using boson sampling, complexity-theoretical studies to reduce the condition will be an important task.




\section*{Acknowledgements}
We thank Owen Howell, Alireza Seif, Roozbeh Bassirian for interesting and fruitful discussions.
C.O. and L.J. acknowledge support from the ARO (W911NF-18-1-0020, W911NF-18-1-0212), ARO MURI (W911NF-16-1-0349), AFOSR MURI (FA9550-19-1-0399), NSF (EFMA-1640959, OMA-1936118, EEC-1941583), NTT Research, and the Packard Foundation (2013-39273).
B.F. acknowledges support from AFOSR (Grant No. YIP FA9550-18-1-0148 and Grant No. FA9550-21-1-0008).
This material is based upon work partially supported by the National Science Foundation under Grant CCF-2044923 (CAREER).
We also acknowledge the University of Chicago’s Research Computing Center for their support of this work.

\appendix
\section{Haar-random Unitary circuit}\label{appendix:Haar-random}
\begin{figure*}[t]
\includegraphics[width=250px]{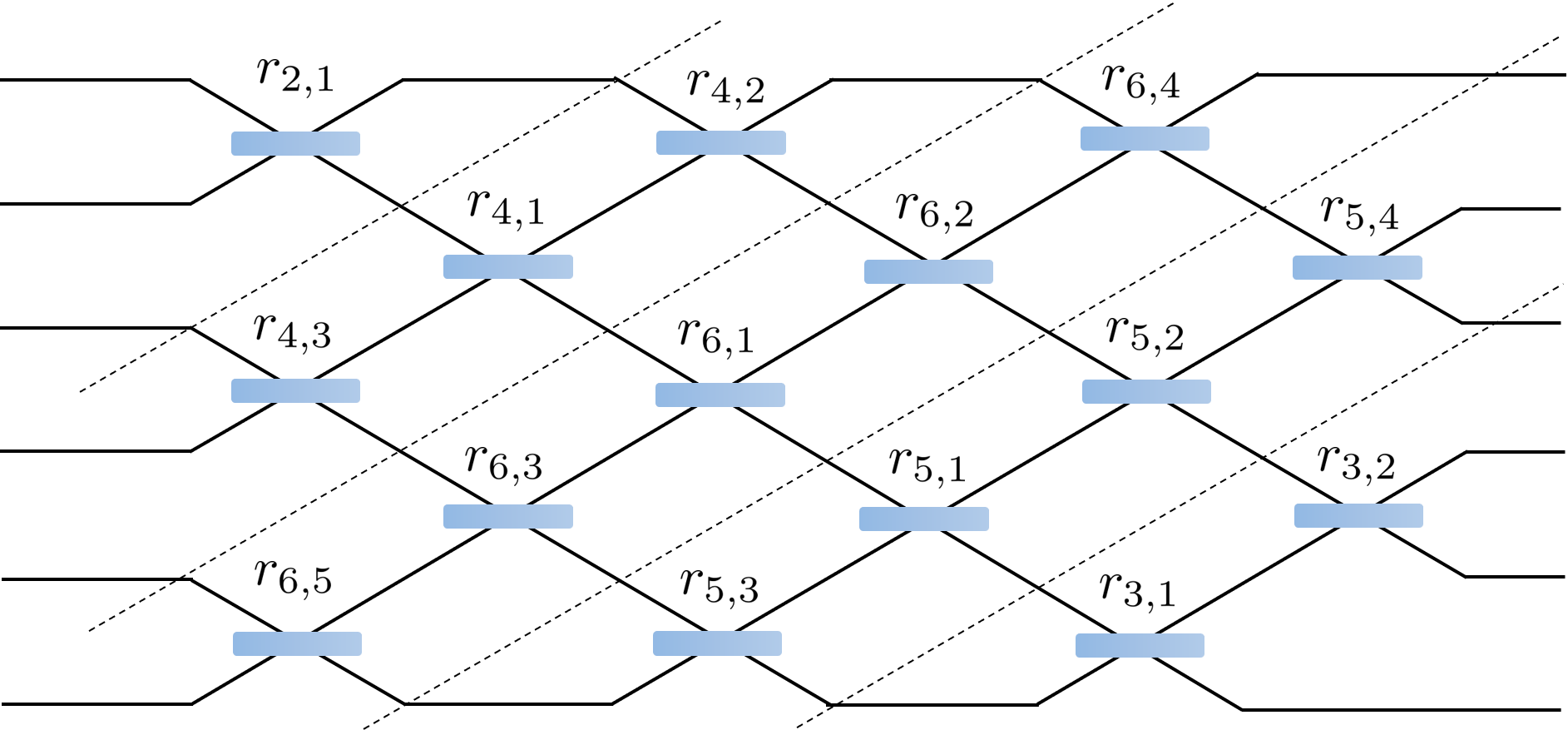}
\caption{Boson sampling circuit with $M=6$. The dashed lines separate different blocks $R_n$, which consists of beam splitters $B_{n,k}$ (See the main text for details.)}
\label{fig:haar}
\end{figure*}

In this Appendix, we present a procedure to implement a Haar-random unitary circuit represented by $U$ for boson sampling.
More details of the procedure can be found in Ref. \cite{russell2017direct}.
Here, the unitary matrix $U$ characterizes the transformation of the mode operators
\begin{align}
    \hat{a}_j\to \sum_{k=1}^M U_{jk} \hat{a}_k.
\end{align}
We assume the number of modes $M$ to be even for simplicity.
First of all, any $M \times M$ unitary matrix $U$ can be written as a product of blocks $R_n$ such that \cite{clements2016optimal}
\begin{align}
U=\prod_{j=1}^{M/2} R_{2j-1} \prod_{i=0}^{M/2-1} R_{M-2i},
\end{align}
where each block $R_{n}$ consists of beam splitters
\begin{align}
R_{n}=\prod_{k\in S_n}B_{n,k}.
\end{align}
Here, $S_n=(s_1,\cdots,s_{n-1})$ is a sequence of $n-1$ indices with odd numbers arranged in descending order and followed by even numbers arranged in ascending order. For example, for $n=4$, $S_4=(3, 1, 2)$, and for $n=5$, $S_5=(3, 1, 2, 4)$, as shown in Fig. \ref{fig:haar}.
The beam splitter of a reflectivity $r$ and a relative phase shift $\phi$ transforms two input modes as
\begin{align}
\begin{pmatrix}
\hat{a}^\dagger \\
\hat{b}^\dagger
\end{pmatrix}
=
\begin{pmatrix}
\sqrt{1-r} & -e^{i\phi}\sqrt{r}  \\ 
e^{-i\phi}\sqrt{r} & \sqrt{1-r}
\end{pmatrix}
\begin{pmatrix}
\hat{a}^\dagger \\
\hat{b}^\dagger
\end{pmatrix},
\end{align}


Note that the circuit depth in Fig. \ref{fig:MPS2} is defined as in Fig. \ref{fig:haar}.

Most importantly, in order to implement Haar-random unitary circuits, the reflectivity $r_{n,i}$ of the beam splitter $B_{n,s_i}$ is sampled from a distribution,
\begin{align}
P_{r_{n,i}}(r)=(n-s_i)(1-r)^{n-s_i-1}.
\end{align}
Each relative phase shift $\phi$ is independently sampled from a uniform distribution on $[0,2\pi)$.
Using this procedure, we have implemented a Haar-random unitary circuit for MPS and MPO simulations.

\begin{widetext}
\section{Matrix Product States (MPS) and Matrix Product Operators (MPO)}
\subsection{Standard MPS and MPO method}\label{appendix:mpsmpo}
In this Appendix, we provide the basic concept of MPS and MPO \cite{vidal2003efficient}.
In principle, any pure quantum states can be represented by an MPS exactly by choosing an appropriate bond dimension $0<\chi\leq d^{\left\lfloor M/2\right\rfloor}$ such that
\begin{align}\label{eq:mps_canonical}
|\psi\rangle&=\sum_{i_1,\cdots,i_M=0}^{d-1}c_{i_1,\cdots,i_M}|i_1,\cdots, i_M\rangle=\sum_{i_1,\cdots,i_M=0}^{d-1}\sum_{\alpha_0,\cdots,\alpha_{M}=0}^{\chi-1}A_{\alpha_0\alpha_1}^{[1]i_1}A_{\alpha_1\alpha_2}^{[2]i_2}\cdots A_{\alpha_{M-1}\alpha_M}^{[M]i_M}|i_1,\cdots, i_M\rangle,
\end{align}
where $d$ is the dimension of a local Hilbert space.
The latter representation is not unique and has a gauge freedom.
Thus, we canonicalize the MPS to fix the gauge as \cite{schollwock2011density}
\begin{align}
|\psi\rangle=\sum_{i_1,\cdots,i_M=0}^{d-1}\sum_{\alpha_0,\cdots,\alpha_{M}=0}^{\chi-1}& \Gamma_{\alpha_0\alpha_1}^{[1]i_1}\lambda_{\alpha_1}^{[1]}\Gamma_{\alpha_1\alpha_2}^{[2]i_2}\lambda_{\alpha_2}^{[2]}\cdots \lambda_{\alpha_{M-1}}^{[M-1]}\Gamma_{\alpha_{M-1}\alpha_{M}}^{[M]i_M}|i_1,\cdots, i_M\rangle.
\end{align}
Here, the vectors $\lambda_{\alpha_k}^{[k]}$ represent the singular values in a spectral decomposition for bipartitions, $|\psi\rangle=\sum_{\alpha_k=0}^{\chi-1}\lambda_{\alpha_k}^{[k]}|\psi_{[1,\cdots, k]}^{\alpha_k}\rangle|\psi_{[k+1,\cdots, M]}^{\alpha_k}\rangle$ with the orthogonality condition on each partition,
\begin{align}
\langle \psi_{[1,\cdots, k]}^{\alpha_k}|\psi_{[1,\cdots, k]}^{\alpha'_k}\rangle=\delta_{\alpha_k,\alpha'_k}, ~\langle \psi_{[k+1,\cdots, M]}^{\alpha_k}|\psi_{[k+1,\cdots, M]}^{\alpha'_k}\rangle=\delta_{\alpha_k,\alpha'_k}.
\end{align}
The singular value vectors $\lambda_{\alpha_k}^{[k]}$ enable one to easily calculate the entanglement entropy (EE) between two partitions.
Also, one of the advantages of the MPS method is that the transformation of a quantum state by a two-site unitary operation acting on $k$ and $k+1$ sites can be efficiently described by updating only the following three relevant tensors with a singular value decomposition,
\begin{align}
\Gamma_{\alpha_{k-1}\alpha_k}^{[k]i_k}, \lambda_{\alpha_k}^{[k]}, \Gamma_{\alpha_k\alpha_{k+1}}^{[k+1]i_{k+1}}.
\end{align}

Specifically, we first write the quantum state in the MPS form as
\begin{align}
|\psi\rangle=\sum_{i_k,i_{k+1}=0}^{d-1}\sum_{\alpha_{k-1}, \alpha_k,\alpha_{k+1}=0}^{\chi-1} \lambda_{\alpha_{k-1}}^{[k-1]}\Gamma_{\alpha_{k-1}\alpha_k}^{[k]i_k}\lambda_{\alpha_{k}}^{[k]}\Gamma_{\alpha_{k}\alpha_{k+1}}^{[k+1]i_{k+1}} \lambda_{\alpha_{k+1}}^{[k+1]}|\psi_{[1,\cdots, k-1]}^{\alpha_{k-1}}\rangle|i_k\rangle|i_{k+1}\rangle |\psi_{[k+2, \cdots, M]}^{\alpha_{k+2}}\rangle.
\end{align}
After the unitary operation acting on $k$ and $k+1$ sites, the state evolves to
\begin{align}
\hat{U}_{k,k+1}|\psi\rangle&=\sum_{i_k,i_{k+1}=0}^{d-1}\sum_{\alpha_{k-1}, \alpha_k,\alpha_{k+1}=0}^{\chi-1} \lambda_{\alpha_{k-1}}^{[k-1]}\Gamma_{\alpha_{k-1}\alpha_k}^{[k]i_k}\lambda_{\alpha_{k}}^{[k]}\Gamma_{\alpha_{k}\alpha_{k+1}}^{[k+1]i_{k+1}} \lambda_{\alpha_{k+1}}^{[k+1]}|\psi_{[1,\cdots, k-1]}^{\alpha_{k-1}}\rangle\hat{U}_{k,k+1}(|i_k\rangle|i_{k+1}\rangle) |\psi_{[k+2, \cdots, M]}^{\alpha_{k+2}}\rangle \\
&=\sum_{j_k,j_{k+1},i_k,i_{k+1}=0}^{d-1}\sum_{\alpha_{k-1}, \alpha_k,\alpha_{k+1}=0}^{\chi-1} \lambda_{\alpha_{k-1}}^{[k-1]}\Gamma_{\alpha_{k-1}\alpha_k}^{[k]i_k}\lambda_{\alpha_{k}}^{[k]}\Gamma_{\alpha_{k}\alpha_{k+1}}^{[k+1]i_{k+1}} \lambda_{\alpha_{k+1}}^{[k+1]}U_{j_k j_{k+1}}^{i_k i_{k+1}}|\psi_{[1,\cdots, k-1]}^{\alpha_{k-1}} \rangle|j_k\rangle|j_{k+1}\rangle |\psi_{[k+2, \cdots, M]}^{\alpha_{k+2}}\rangle \\
&=\sum_{j_k,j_{k+1}=0}^{d-1}\sum_{\alpha_{k-1},\alpha_{k+1}=0}^{\chi-1} \Theta_{\alpha_{k-1},\alpha_{k+1}}^{j_k,j_{k+1}}|\psi_{[1,\cdots, k-1]}^{\alpha_{k-1}} \rangle|j_k\rangle|j_{k+1}\rangle |\psi_{[k+2, \cdots, M]}^{\alpha_{k+2}}\rangle,
\end{align}
where we defined
\begin{align}
\Theta^{j_k j_{k+1}}_{\alpha_{k-1}\alpha_{k+1}}&=\sum_{i_k,i_{k+1}=0}^{d-1}\sum_{\alpha_k=0}^{\chi-1} U_{i_k i_{k+1}}^{j_k j_{k+1}}\lambda_{\alpha_{k-1}}^{[k-1]}\Gamma_{\alpha_{k-1}\alpha_k}^{[k]i_k}\lambda_{\alpha_k}^{[k]}\Gamma_{\alpha_k\alpha_{k+1}}^{[k+1]i_{k+1}}\lambda_{\alpha_{k+1}}^{[k+1]},~~~~ U^{j_k j_{k+1}}_{i_k i_{k+1}}=\langle j_k,j_{k+1}|\hat{U}_{k,k+1}|i_k,i_{k+1}\rangle.
\end{align}
Note that the complexity of obtaining $\Theta$ is $O(d^4\chi^3)$.
We now perform singular value decomposition of $\Theta$ to recover the MPS representation of the evolved state,
\begin{align}
\Theta^{j_k j_{k+1}}_{\alpha_{k-1}\alpha_{k+1}}=\sum_{\beta_k=0}^{d\chi-1}V_{(j_k, \alpha_{k-1}), \beta_k} \tilde{\lambda}^{[k]}_{\beta_k} W_{\beta_k,(j_{k+1},\alpha_{k+1})}  \approx \sum_{\alpha_k=0}^{\chi-1}\lambda_{\alpha_{k-1}}^{[k-1]}\tilde{\Gamma}_{\alpha_{k-1}\alpha_k}^{[k]i_k}\tilde{\lambda}_{\alpha_k}^{[k]}\tilde{\Gamma}_{\alpha_k\alpha_{k+1}}^{[k+1]i_{k+1}}\lambda_{\alpha_{k+1}}^{[k+1]}.
\end{align}
In the approximation, we keep the largest $\chi$  singular values $\tilde{\lambda}_{\beta_k}^{[k]}$.
Also, we defined
\begin{align}
\tilde{\Gamma}_{\alpha_{k-1}\alpha_k}^{[k]i_k}=V_{(j_k,\alpha_{k-1}),\beta_k}/\lambda_{\alpha_{k-1}}^{[k-1]}, ~~~ \tilde{\Gamma}_{\alpha_{k}\alpha_{k+1}}^{[k+1]i_{k+1}}=W_{\beta_k,(j_{k+1},\alpha_{k+1})}/\lambda_{\alpha_{k+1}}^{[k+1]}.
\end{align}
Thus, we obtain the MPS representation after two-site unitary operation.
We note that one can increase the accuracy of the simulation by transforming the MPO into an orthogonal form by performing QR decomposition after the truncation \cite{zhou2020what}.

Consequently, since the dominant time cost comes from matrix multiplications and singular value decomposition, the computational time cost for a two-site unitary update is $T=O(d^4\chi^3)$ which accounts for matrix multiplications and singular value decomposition of a $d\chi \times d\chi$ matrix.
It implies that the computational cost for MPS simulations depends on the bond dimension $\chi$ we choose.
As a result, the computational cost to implement MPS simulation for boson sampling circuits is given by
\begin{align}
T=O(D M d^4\chi^3),
\end{align}
where $D$ and $M$ accounts for the number of beam splitter layers and the number of two-site unitary operators in each layer, respectively.

From now on, let us consider MPO representation and a two-site unitary operator on $k$ and $k+1$ sites to describe mixed states.
Similarly, the MPO representation can be updated easily.
We first vectorize a density matrix $\hat{\rho}$ as
\begin{align}
\hat{\rho}&=\sum_{i_1,i_1',\cdots, i_M, i_M'=0}^{d-1}\rho_{i_1,i_1',\cdots, i_M,i'_M}|i_1,\cdots, i_M\rangle\langle i_1',\cdots, i_M'| \nonumber \\
\to |\hat{\rho}\rangle\rangle
&=\sum_{i_1,\bar{i}_1',\cdots,i_M,\bar{i}_M'=0}^{d-1}\sum_{\alpha_0,\cdots,\alpha_{M}=0}^{\chi-1}\Gamma_{\alpha_0\alpha_1}^{[1]i_1\bar{i}_1'}\lambda_{\alpha_1}^{[1]}\Gamma_{\alpha_1\alpha_2}^{[2]i_2\bar{i}_2}\lambda_{\alpha_2}^{[2]} \cdots \lambda_{\alpha_{M-1}}^{[M-1]}\Gamma_{\alpha_{M-1}\alpha_M}^{[M]i_M\bar{i}_M'}|i_1,\bar{i}_1',\cdots, i_M, \bar{i}_M'\rangle\rangle.
\end{align}
The MPO can be rewritten as
\begin{align}
|\hat{\rho}\rangle\rangle=\sum_{I_k,I_{k+1}=0}^{d^2-1}\sum_{\alpha_{k-1}, \alpha_k,\alpha_{k+1}=0}^{\chi-1} \lambda_{\alpha_{k-1}}^{[k-1]}\Gamma_{\alpha_{k-1}\alpha_k}^{[k]I_k}\lambda_{\alpha_{k}}^{[k]}\Gamma_{\alpha_{k}\alpha_{k+1}}^{[k+1]I_{k+1}} \lambda_{\alpha_{k+1}}^{[k+1]}|\psi_{[1,\cdots, k-1]}^{\alpha_{k-1}}\rangle\rangle|I_k\rangle\rangle|I_{k+1}\rangle\rangle |\psi_{[k+2, \cdots, M]}^{\alpha_{k+2}}\rangle\rangle,
\end{align}
where $I_k\equiv i_k+d\bar{i}_{k}$ and $|I_k\rangle\rangle\equiv|i_k,\bar{i}_k\rangle\rangle$.
After the two-site unitary operation, the vectorized density matrix is transformed to
\begin{align}
|\hat{\rho}'\rangle\rangle&=\sum_{J_k,J_{k+1},I_k,I_{k+1}=0}^{d^2-1}\sum_{\alpha_{k-1}, \alpha_k,\alpha_{k+1}=0}^{\chi-1} \lambda_{\alpha_{k-1}}^{[k-1]}\Gamma_{\alpha_{k-1}\alpha_k}^{[k]I_k}\lambda_{\alpha_{k}}^{[k]}\Gamma_{\alpha_{k}\alpha_{k+1}}^{[k+1]I_{k+1}} \lambda_{\alpha_{k+1}}^{[k+1]}\mathcal{U}_{J_k J_{k+1}}^{I_k I_{k+1}}|\psi_{[1,\cdots, k-1]}^{\alpha_{k-1}} \rangle\rangle|J_k\rangle\rangle|J_{k+1}\rangle\rangle |\psi_{[k+2, \cdots, M]}^{\alpha_{k+2}}\rangle\rangle \\
&=\sum_{J_k,J_{k+1}=0}^{d^2-1}\sum_{\alpha_{k-1},\alpha_{k+1}=0}^{\chi-1} \Theta_{\alpha_{k-1},\alpha_{k+1}}^{J_k,J_{k+1}}|\psi_{[1,\cdots, k-1]}^{\alpha_{k-1}} \rangle\rangle|J_k\rangle\rangle|J_{k+1}\rangle\rangle |\psi_{[k+2, \cdots, M]}^{\alpha_{k+2}}\rangle\rangle,
\end{align}
where
\begin{align}
\mathcal{U}_{J_k J_{k+1}}^{I_k I_{k+1}}&\equiv \langle j_k j_{k+1}|\hat{U}|i_k i_{k+1}\rangle\langle \bar{i}_{k}\bar{i}_{k+1}|\hat{U}^\dagger |\bar{j}_k \bar{j}_{k+1}\rangle, \\ 
\Theta^{J_k J_{k+1}}_{\alpha_{k-1}\alpha_{k+1}}&\equiv \sum_{I_k,I_{k+1}=0}^{d^2-1}\sum_{\alpha_k=0}^{\chi-1} \mathcal{U}_{I_k I_{k+1}}^{J_k J_{k+1}}\lambda_{\alpha_{k-1}}^{[k-1]}\Gamma_{\alpha_{k-1}\alpha_k}^{[k]I_k}\lambda_{\alpha_k}^{[k]}\Gamma_{\alpha_k\alpha_{k+1}}^{[k+1]I_{k+1}}\lambda_{\alpha_{k+1}}^{[k+1]}.
\end{align}
The time complexity of obtaining $\Theta$ is given by $O(d^8\chi^3)$
Again, we perform singular value decomposition and keep the $\chi$ largest singular values only,
\begin{align}
\Theta^{J_k J_{k+1}}_{\alpha_{k-1}\alpha_{k+1}}=\sum_{I_k,I_{k+1}=0}^{d^2-1}\sum_{\beta_k=0}^{d^2\chi-1}\tilde{V}_{(J_k, \alpha_{k-1}), \beta_k} \tilde{\lambda}^{[k]}_{\beta_k} \tilde{W}_{\beta_k,(J_{k+1},\alpha_{k+1})}
\approx \sum_{\alpha_k=0}^{\chi-1}\lambda_{\alpha_{k-1}}^{[k-1]}\tilde{\Gamma}_{\alpha_{k-1}\alpha_k}^{[k]I_k}\tilde{\lambda}_{\alpha_k}^{[k]}\tilde{\Gamma}_{\alpha_k\alpha_{k+1}}^{[k+1]I_{k+1}}\lambda_{\alpha_{k+1}}^{[k+1]},
\end{align}
which is the updated MPO representation after the unitary operation.
Thus, the total computational time cost for boson sampling circuits is given by
\begin{align}
T=O(DM d^8\chi^3).
\end{align}

One can easily check that normalization of $\sum_{\alpha_k=0}^{\chi-1}\lambda_{\alpha_k}^2$ is conserved for unitary updates if there is no truncation error.
Note that for an arbitrary $n$ by $m$ matrix $A$, $\sum_{\alpha=1}^{\text{min}(n,m)}\lambda_{\alpha}^2=\sum_{i,j=1}^{n,m}|A_{i,j}|^2$, where $\lambda_{\alpha}$ is singular values.
Using the unitarity $\hat{U}\hat{U}^\dagger=\hat{U}^\dagger\hat{U}=\mathbb{1}$,
\begin{align}
    \sum_{J_k,J_{k+1}=0}^{d^2-1}\mathcal{U}_{I_k I_{k+1}}^{J_k J_{k+1}}\mathcal{U}_{I'_k I'_{k+1}}^{*J_k J_{k+1}}
    &=\sum_{J_k,J_{k+1}=0}^{d^2-1}\langle j_k j_{k+1}|\hat{U}|i_k i_{k+1}\rangle\langle \bar{i}_{k}\bar{i}_{k+1}|\hat{U}^\dagger |\bar{j}_k \bar{j}_{k+1}\rangle \langle i'_k i'_{k+1}|\hat{U}^\dagger|j_k j_{k+1}\rangle\langle \bar{j}_k \bar{j}_{k+1}|\hat{U}^\dagger |\bar{i}_{k}\bar{i}_{k+1}\rangle \nonumber \\ 
    &=\delta_{i_k,i'_k}\delta_{i_{k+1},i'_{k+1}}\delta_{\bar{i}_k,\bar{i}'_k}\delta_{\bar{i}_{k+1},\bar{i}'_{k+1}}=\delta_{I_k,I_k'}\delta_{I_{k+1},I'_{k+1}},
\end{align}
and for some unitary matrix $V$ and $W$,
\begin{align}
    \sum_{\alpha_{k-1}=0}^{\chi-1}\sum_{I_k=0}^{d^2-1}\left(\lambda_{\alpha_{k-1}}^{[k-1]}\right)^2 \Gamma_{\alpha_{k-1}\alpha_k}^{[k]I_k}\Gamma_{\alpha_{k-1}\alpha_k'}^{*[k]I_k}&=\sum_{\alpha_{k-1}=0}^{\chi-1}\sum_{I_k=0}^{d^2-1} V_{(I_k,\alpha_{k-1}),\alpha_k}V^*_{(I_k,\alpha_{k-1}),\alpha_k'}=\delta_{\alpha_k,\alpha_k'}, \\
    \sum_{\alpha_{k+1}=0}^{\chi-1}\sum_{I_{k+1}=0}^{d^2-1}\left(\lambda_{\alpha_{k+1}}^{[k]}\right)^2 \Gamma_{\alpha_k\alpha_{k+1}}^{[k]I_{k+1}}\Gamma_{\alpha_k'\alpha_{k+1}}^{*[k]I_{k+1}}&=\sum_{\alpha_{k+1}=0}^{\chi-1}\sum_{I_{k+1}=0}^{d^2-1} W_{\alpha_k,(I_{k+1},\alpha_{k+1})}W^*_{\alpha_k',(I_{k+1},\alpha_{k+1})}=\delta_{\alpha_k,\alpha_k'},
\end{align}
one can show that 
\begin{align}
    \sum_{\beta_k=0}^{\chi-1}\tilde{\lambda}_{\beta_k}^2=\sum_{J_k,J_{k+1}=0}^{d^2-1}\sum_{\alpha_{k-1},\alpha_{k+1}=0}^{\chi-1}|\Theta_{\alpha_{k-1},\alpha_{k+1}}^{J_k,J_{k+1}}|^2=\sum_{\alpha_k=0}^{\chi-1}\lambda_{\alpha_k}^2.
\end{align}

\subsection{MPS / MPO simulation using U(1) symmetry} \label{appendix:simul}
In this Appendix, we introduce a method to simulate boson sampling using an MPS representation with U(1) symmetry, which can be used to improve an MPS simulation more efficiently \cite{singh2011tensor, guo2019matrix, huang2019simulating}.
Basically, we enforce the global U(1) symmetry by introducing a charge vector $c^{[k]}_{\alpha_k}$ on each bond index $\alpha_k$.
Here, a charge $c_{\alpha_k}^{[k]}$ accounts for the photon number occupied by the right-hand side of the bipartition of $[1\cdots k]:[(k+1)\cdots M]$ for a given bond index $\alpha_k$.
Since the total photon number $N$ is fixed in the system, the charges on each end are set as $c^{[0]}_{\alpha_0=0}=N$ and $c^{[M]}_{\alpha_M=0}=0$.
For example, consider the following state:
\begin{align}
    |\psi\rangle=|1100\rangle,
\end{align}
which is an initial state when $M=4$ and $N=2$.
Since the total photon number is $N=2$, we initialize $c_0^{[0]}=2$ and $c_0^{[4]}=0$.
Charge vectors for different bipartitions can be easily determined by counting how many photons the right-hand side of a bipartition occupies.
For bipartition $[1]:[2,3,4]$, the charge vector becomes $c^{[1]}_0=1$ because the partition $[2,3,4]$ is occupied by a single photon, and for bipartition $[1,2]:[3,4]$ and $[1,2,3]:[4]$, the charge vectors become $c^{[2]}_0=0$ and $c^{[3]}_0=0$ because there is no photon for the right-hand side partition.
As a result of the charge conservation, only tensor elements $\Gamma_{\alpha_k \alpha_{k+1}}^{[k]i_{k+1}}$ that satisfy the constraint $c^{[k]}_{\alpha_k}-c^{[k+1]}_{\alpha_{k+1}}=i_{k+1}$ are non-vanishing \cite{singh2011tensor, guo2019matrix, huang2019simulating}.
Thus, each tensor $\Gamma_{\alpha_k\alpha_{k+1}}^{[k]i_{k+1}}$ for different $i_{k+1}$ is compressed by a tensor $\Gamma_{\alpha_k\alpha_{k+1}}^{[k]}$ with charge vectors $c_{\alpha_k}^{[k]}$ and $c_{\alpha_{k+1}}^{[k+1]}$.
Consequently, in contrast to a typical MPS without U(1) symmetry where tensors $\{\Gamma_{\alpha_k,\alpha_{k+1}}^{[k]i_k},\lambda^{[k]}_{\alpha_k}\}$ constitute an MPS, here, charge vector $c^{[k]}_{\alpha_k}$ has to be added as $\{\Gamma_{\alpha_k,\alpha_{k+1}}^{[k]},\lambda^{[k]}_{\alpha_k},c_{\alpha_k}^{[k]}\}$ to fully describe a given quantum state.

Notably, the memory usage of an MPS simulation is significantly reduced because local indices $i_{k+1}$ are dropped.
Specifically, whereas an original tensor $\Gamma_{\alpha_k,\alpha_{k+1}}^{[k]i_{k+1}}$ without U(1) symmetry requires $O(d\chi^2)$ memories for the local index $d$ and two bond indices $\chi$, since local indices are dropped, we only need $O(\chi^2)$ for a single tensor and $O(\chi)$ for a charge vector.
Thus, taking into account singular value vectors and charge vectors, a total memory cost is given by $O(M\chi^2+(M-1)\chi+(M+1)\chi)$, which is significantly reduced from a memory cost $O(Md \chi^2+(M-1)\chi)$ required without U(1) symmetry.
In principle, the probability amplitude $c_{i_1,\cdots, i_M}$ in Eq.~\eqref{eq:mps_canonical} can be reproduced by
\begin{align}
c_{i_1,\cdots, i_M}&=\sum_{\alpha_0,\cdots,\alpha_{M}=0}^{\chi-1}\Gamma_{\alpha_0\alpha_1}^{[1]}\lambda_{\alpha_1}^{[1]}\Gamma_{\alpha_1\alpha_2}^{[2]}\cdots \lambda_{\alpha_{M-1}}^{[M-1]}\Gamma_{\alpha_{M-1}\alpha_{M}}^{[M]}\prod_{k=1}^{M} \delta(c_{\alpha_{k-1}}^{[k-1]}-c_{\alpha_{k}}^{[k]}-i_{k}),
\end{align}
where the delta function indicates the constraint $c_{\alpha_{k-1}}^{[k-1]}-c_{\alpha_{k}}^{[k]}=i_{k}$ for $1\leq k \leq M$. The delta function is defined as $\delta(0)=1$ and zero otherwise.

In addition, a computational time cost of a canonical update for a two-site unitary can also be reduced as follows.
Let us consider a two-site unitary acting on $k$ and $k+1$ sites,
where the relevant tensors for the update are
\begin{align}
\lambda_{\alpha_{k-1}}^{[k-1]}, \Gamma^{[k]}_{\alpha_{k-1}\alpha_k}, \lambda_{\alpha_k}^{[k]}, \Gamma^{[k+1]}_{\alpha_{k}\alpha_{k+1}}, \lambda_{\alpha_{k+1}}^{[k+1]}.
\end{align}
For all $0 \leq c^{[k]} \leq N$, we multiply the unitary matrix and obtain
\begin{align}\label{eq:multi_uni}
\Theta_{\alpha_{k-1},\alpha_{k+1}}^{i_k,i_{k+1}}(c^{[k]})
&=\sum_{j_k,j_{k+1}=0}^{d-1}\sum_{\alpha_k=0}^{\chi-1}U^{i_k,i_{k+1}}_{j_k,j_{k+1}}\lambda_{\alpha_{k-1}}^{[k-1]}\Gamma^{[k]}_{\alpha_{k-1}\alpha_k}\lambda_{\alpha_k}^{[k]}\Gamma^{[k+1]}_{\alpha_{k}\alpha_{k+1}}\lambda_{\alpha_{k+1}}^{[k+1]} \\ 
&\times 
\delta(c_{\alpha_{k-1}}^{[k-1]}-c_{\alpha_k}^{[k]}-j_{k})\delta(c_{\alpha_k}^{[k]}-c^{[k+1]}_{\alpha_{k+1}}-j_{k+1})
\delta(c_{\alpha_{k-1}}^{[k-1]}-c^{[k]}-i_{k})\delta(c^{[k]}-c^{[k+1]}_{\alpha_{k+1}}-i_{k+1}),
\end{align}
where the first two delta functions correspond to the constraints of the input photon numbers and the last two delta functions to the constraints of the output photon numbers.
Thus, the complexity of computing $\Theta$ is given as $O(d^5\chi^3)$.
We note that such a scaling is conservative in the sense that the bond dimension $\chi$ is partitioned according to different charges so that a partitioned bond dimension is much smaller than $\chi$ and the scaling is smaller in practice.
Hence, U(1) symmetry highly decreases the computational cost in practice.
Since the scaling from $d$ is polynomial, the bond dimension $\chi$ is the important parameter that determines if an efficient simulation is possible.

In order to obtain the updated tensors, we perform singular value decompositions,
\begin{align}
\Theta_{\alpha_{k-1},\alpha_{k+1}}^{i_k,i_{k+1}}(c^{[k]})=\sum_{\beta_k} V_{(i_k,\alpha_{k-1}),\beta_k}\tilde{\lambda}_{\beta_k}^{[k]}W_{\beta_k,(i_{k+1},\alpha_{k+1})},
\end{align}
where we assign the charge $c^{[k]}$ for each $\beta_k$.
After iterating the same procedure for all $0\leq c^{[k]}\leq N$, we update a singular value vector by choosing the largest $\chi$ singular values only among all singular values of $\beta_k$ and relabeling them as $0\leq \alpha_k\leq\chi-1$.
A charge vector $c_{\alpha_k}^{[k]}$ is updated by $c^{[k]}$ that corresponds to $\alpha_k$. 
Finally, tensors are accordingly updated,
\begin{align}
\Gamma_{\alpha_{k-1}\alpha_k}^{[k]}=V_{(i_k,\alpha_{k-1}),\alpha_k}/\lambda_{\alpha_{k-1}}^{[k-1]},~~~
\Gamma_{\alpha_{k}\alpha_{k+1}}^{[k+1]}=W_{\alpha_k,(i_{k+1},\alpha_{k+1})}/\lambda_{\alpha_{k+1}}^{[k+1]}.
\end{align}
For example, let us consider a beam-splitter operation on the first and second mode, which transforms a state as follows:
\begin{align}
    |100\rangle\rightarrow \frac{1}{\sqrt{2}}(|100\rangle+|010\rangle).
\end{align}
After multiplying the unitary matrix as Eq.~\eqref{eq:multi_uni} and performing a singular value decomposition, we obtain tensors corresponding to $|100\rangle/\sqrt{2}$ for $c^{[1]}=0$ and tensors corresponding to $|010\rangle/\sqrt{2}$ for $c^{[1]}=1$ due to the charge constraints in Eq.~\eqref{eq:multi_uni}.
As a result, the elements of the initial charge vector $c_0^{[1]}=0$ are updated to $c_0^{[1]}=0$ and $c_1^{[1]}=1$ after the beam-splitter.
Since a singular value decomposition is performed for different charges, the matrix size for each singular value decomposition is significantly reduced, which results in a reduction of the computation time.

Even when a given quantum state is a superposition of different photon number states, one can still use the U(1) symmetry in such a way that
a charge on the left edge $c^{[1]}_{\alpha_0}$ has different conserved charges \cite{guo2019matrix}, which will be elaborated below.

We can employ U(1) symmetry for an MPO simulation with a slight modification of an MPS simulation \cite{guo2019matrix}.
As a byproduct of vectorization, we have two different conserved charges corresponding to indices $i_k$'s and $\bar{i}_k'$'s.
Thus, charge vectors now save two different charges $(n, m)$.
Since initial states for lossy boson sampling do not have a definite photon number as shown in Eq.~\eqref{ini_state}, 
a straightforward extension of MPS simulation with U(1) symmetry to MPO simulation is to decompose an initial state depending on the total charge and execute unitary updates separately.
In other words, we constitute $N+1$ different MPOs having a different total charge by setting conserved charges $(n,n)$ on the left end $c^{[0]}_{\alpha_0=0}$ for the $n$th MPO and on the right end as $c_{\alpha_M=0}^{[M]}=(0, 0)$.
For example, consider the following bipartite state 
\begin{align}
\frac{1}{2}(|00\rangle\rangle+|11\rangle\rangle)\otimes \frac{1}{2}(|00\rangle\rangle+|11\rangle\rangle).
\end{align}
Without using U(1) symmetry, the singular values of the state are given by $\lambda^{[1]}_{\alpha_1=0}=1$ and $\lambda^{[1]}_{\alpha_1>0}=0$ because it is a product state.
However, since we are using U(1) symmetry, the state is decomposed as
\begin{align}
\frac{1}{4}|00\rangle\rangle\otimes|00\rangle\rangle, ~~~\frac{1}{4}|11\rangle\rangle\otimes|11\rangle\rangle, ~~~\frac{1}{4}(|00\rangle\rangle\otimes|11\rangle\rangle+|11\rangle\rangle\otimes|00\rangle\rangle).
\end{align}
Here, conserved charges for each element are the sum of the first elements in the vector form and that of the second elements.
In this example, we have three subspaces whose conserved charges are $(0,0), (2,2)$ and $(1, 1)$, respectively.
Thus, one may constitute MPOs for each conserved charge separately.
In addition, if one wants to simulate a post-selected total photon number as done in Ref. \cite{wang2019boson}, this procedure can be used to simulate the dynamics by selecting a desired total charge.

However, one can improve the simulation more efficiently by combining all the MPOs by assigning charges on the left end as $c^{[0]}_{\alpha_0=n}=(n,n)$ for $0\leq n \leq N$.
The unitary update of the latter method is more consistent since the truncation of singular values is performed at the same time.
For this reason, we use the latter method for MPO simulation.
The procedure of unitary updates is similar to MPS simulation.
The only difference is that the charge vector now consists of two components, so we iterate $d^2$ times singular value decompositions.
Thus, the time cost to compute $\Theta$ in Eq.~\eqref{eq:multi_uni} for MPO is given by $O(d^{10}\chi^3)$.
Again, this scaling is conservative and the practical complexity is much smaller because the bond dimension is partitioned according to charges so that a bond dimension in each partition is reduced.
Thus, in practice, U(1) symmetry highly reduces the computational cost.
Also, since the scaling from $d$ is polynomial, the bond dimension $\chi$ is the parameter that determines if an efficient simulation is possible.
Memory saving from U(1) symmetry is more significant in MPO simulation because the local indices up to $d^2$ can be dropped.

As a remark, in the case of non-uniform loss \cite{brod2020classical}, we cannot simplify the problem by merging all loss channels as we did for uniform loss because non-uniform loss channels do not commute with beam splitters in general.
Therefore, one needs to update an MPO by a completely positive trace-preserving map for a loss-channel for each step, which requires more computational time \cite{noh2020efficient}. 
In addition, we may not be able to take advantage from symmetry because loss-channels do not preserve global U(1) symmetry.

\subsection{Computing outcome probabilities and sampling outcomes from MPS / MPO}\label{appendix:sample}
Now, we present how to compute outcome probabilities and sample outcomes according to the probability distribution using MPS and MPO \cite{noh2020efficient}.
The probability of obtaining a given outcome $\vec{n}$ is written as
\begin{align}\label{born1}
P_{|\psi\rangle}(\vec{n})\equiv |\langle \psi|\vec{n}\rangle|^2,
\end{align}
for pure states, and
\begin{align}\label{born2}
P_{\hat{\rho}}(\vec{n})\equiv \text{Tr}[\hat{\rho}|\vec{n}\rangle\langle \vec{n}]]=\langle\langle \vec{n}|\hat{\rho}\rangle\rangle,
\end{align}
for mixed states. Here, $|\vec{n}\rangle=|n_1,\cdots, n_M\rangle$ corresponds to the outcome $\vec{n}$.
First of all, a marginal probability can be efficiently computed. For example, a probability to detect $(n_1,\cdots,n_l)$ on the first $l$ modes is given by
\begin{align}\label{mps:marginal}
P_{\hat{\rho}}^{[1,\cdots, l]}(n_1,\cdots, n_l)=\bigg|\sum^{\chi-1}_{\alpha_0,\cdots,\alpha_{M}=0}\Gamma_{\alpha_0\alpha_1}^{[1]}\lambda_{\alpha_1}^{[1]} \cdots \Gamma_{\alpha_{M-1}\alpha_M}^{[M]} \prod_{k=1}^{l}\delta(c^{[k-1]}_{\alpha_{k-1}}-c^{[k]}_{\alpha_{k}}-n_{k})\bigg|^2
\end{align}
for an MPS and 
\begin{align}\label{mpo:marginal}
P_{\hat{\rho}}^{[1,\cdots, l]}(n_1,\cdots, n_l)=\sum_{\alpha_0,\cdots,\alpha_{M}=0}^{\chi-1}\Gamma_{\alpha_0\alpha_1}^{[1]}\lambda_{\alpha_1}^{[1]}\cdots \Gamma_{\alpha_{M-1}\alpha_M}^{[M]}\prod_{k=1}^l \delta(c^{[k-1]}_{\alpha_{k-1}}-c^{[k]}_{\alpha_k}-(n_k, n_k))
\end{align}
for an MPO.
Using the above equations, one can easily find that an outcome probability for $\vec{n}$ can be obtained by setting $l=M$.

Now, we present a procedure to sample an outcome from MPS / MPO representation.
First, one computes a marginal probability to detect $n_1$ at the first mode $P^{[1]}(n_1)$ using Eq.~\eqref{mps:marginal} or Eq.~\eqref{mpo:marginal}.
After obtaining the first outcome $n_1^*$, we sample $n_2$ from the conditional probability distribution which can be efficiently found by using
\begin{align}
P^{[2|1]}(n_2|n_1^*)=\frac{P^{[1,2]}(n_1^*,n_2)}{P^{[1]}(n_1^*)}.
\end{align}
We sample the remaining measurement outcomes following the same rule as
\begin{align}
P^{[k+1|1\cdots k]}(n_{k+1}|n_1^*,\cdots,n_k^*)=\frac{P^{[1\cdots (k+1)]}(n_1^*,\cdots, n_{k+1})}{P^{[1\cdots k]}(n_1^*,\cdots, n_k^*)}.
\end{align}
We finally obtain $\vec{n}=(n_1^*,\cdots,n_M^*)$ that follows Born's rule as in Eqs.~\eqref{born1} and \eqref{born2}, which can be efficiently performed.

\section{Entanglement entropy of Matrix Product states and Matrix Product operators}\label{appendix:entropy}
Let us consider a beam splitter array, which transforms the creation operators of input modes $\hat{a}^\dagger_j$ into the creation operators of output modes $\hat{b}^\dagger_j$ as
\begin{align}
    \hat{a}_j^\dagger \to \hat{b}_j^\dagger=\hat{U}^\dagger\hat{a}_j^\dagger\hat{U}=\sum_{k=1}^M U_{jk} \hat{a}_k^\dagger.
\end{align}
To obtain entanglement entropy between partitions $[1,\cdots, l]$ and $[(l+1),\cdots,M]$, we rewrite the output mode operators as
\begin{align}
    \hat{b}^\dagger_j=\cos\theta_j\hat{B}^\dagger_{\text{up},j}+\sin\theta_j\hat{B}^\dagger_{\text{down},j},
\end{align}
where we defined normalized creation operators for each partition as
\begin{align}
    \cos\theta_{j}\hat{B}^\dagger_{\text{up},j}=\sum_{k=1}^l U_{jk}\hat{a}_k^\dagger,~~~\sin\theta_{j}\hat{B}^\dagger_{\text{down},j}=\sum_{k=l+1}^M U_{jk}\hat{a}_k^\dagger,
\end{align}
and their normalization as
\begin{align}
    \cos^2\theta_{j}\equiv \frac{\sum_{k=1}^l|U_{jk}|^2}{\sum_{k=1}^M|U_{jk}|^2}=\sum_{k=1}^l|U_{jk}|^2,~~~\sin^2\theta_{j}\equiv \frac{\sum_{k=l+1}^M|U_{jk}|^2}{\sum_{k=1}^M|U_{jk}|^2}=\sum_{k=l+1}^M|U_{jk}|^2.
\end{align}
Note that assuming collision-free cases $M\geq N^2$ \cite{aaronson2011computational}, the creation operators $\hat{B}^\dagger_{\text{up},j}, \hat{B}^\dagger_{\text{down},j}$ satisfy the canonical commutation relations, 
\begin{align}
    [\hat{B}_{\text{up},j},\hat{B}_{\text{up},k}^\dagger]=\delta_{jk},~~~[\hat{B}_{\text{down},j},\hat{B}_{\text{down},k}^\dagger]=\delta_{jk},~~~ [\hat{B}_{\text{up},j},\hat{B}_{\text{down},k}]=0,~~~ [\hat{B}_{\text{up},j},\hat{B}_{\text{down},k}^\dagger]=0.
\end{align}
For typical random beam splitter arrays with a large number of modes $M\gg 1$, we will have $\cos^2{\theta_j}\approx \sin^2{\theta_j}\approx 1/2$ for $l=M/2$.

Let us first consider the input state occupied by $N_j$ photons for $j$th modes,
\begin{align}
    |\psi_\text{in}\rangle&=\left(\prod_{j=1}^{M} \frac{\hat{a}^{\dagger N_j}_j}{\sqrt{N_j!}}\right)|0\rangle\to |\psi_\text{out}\rangle=\prod_{j=1}^M\frac{1}{\sqrt{N_j!}}\left( \cos\theta_j\hat{B}^\dagger_{\text{up},j}+\sin\theta_j\hat{B}^\dagger_{\text{down},j}\right)^{N_j}|0\rangle \\ 
    &=\otimes_{j=1}^M\left(\sum_{k_j=0}^{N_j} \sqrt{\frac{k_j!(N_j-k_j)!}{N_j!}}\binom{N_j}{k_j}\cos^{k_j}\theta_j\sin^{N_j-k_j}\theta_j |k_j\rangle_{\text{up},j} |N_j-k_j\rangle_{\text{down},j}\right).
\end{align}
The reduced density matrix of the output state for a partition $[1,\cdots, l]$ is then written as
\begin{align}
    \hat{\rho}_\text{up}\equiv\text{Tr}_{[(l+1),\cdots,M]}|\psi_\text{out}\rangle\langle\psi_\text{out}|=\otimes_{j=1}^M \left(\sum_{k_j=0}^{N_j}\binom{N_j}{k_j}\cos^{2k_j}\theta_j \sin^{2(N_j-k_j)}\theta_j|k_j\rangle\langle k_j|_{\text{up},j}\right).
\end{align}
Now, assuming $M\gg 1$, we can approximate $\cos^2{\theta_j}\approx \sin^2{\theta_j}\approx 1/2$ for $l=M/2$, and thus the density matrix becomes
\begin{align}
    \hat{\rho}_\text{up}\approx \otimes_{j=1}^M \left(\sum_{k_j=0}^{N_j}\frac{1}{2^{N_j}} \binom{N_j}{k_j}|k_j\rangle\langle k_j|_{\text{up},j}\right),
\end{align}
which is a product of states whose eigenvalues follow a binomial distributions.
Thus, the entanglement entropy is given by the sum of the entanglement entropy of each state.
It indicates that if we add more modes occupied by at least a single photon, the entanglement entropy increases linearly.
In contrast, if we increase the number of photon in each mode and assume that $N_j\gg 1$, then 
the entanglement entropy can be approximated as
\begin{align}
    S(\hat{\rho}_\text{up})\approx \frac{1}{2}\sum_{j=1}^{M}\log_2\left(\frac{\pi e N_j}{2}\right).
\end{align}
where we have used a Gaussian approximation of binomial distribution.
Thus, the entanglement entropy increases logarithmically of $N_j$, which suggests that the MPS simulation can be efficiently performed for $N_j$.
Here, note that when $N_j$ is zero for some modes, we treat the entropy to be zero for the modes in the summation.

Particularly, let us first consider the input state of the standard boson sampling, where $N_j=1$ for $1\leq j \leq N$ and otherwise $N_j=0$.
One can immediately see that for large $M$, the reduced density matrix is written as
\begin{align}
    \hat{\rho}_\text{up}\approx \otimes_{j=1}^N \frac{1}{2}\left(|0\rangle\langle 0|_{\text{up},j}+|1\rangle\langle 1|_{\text{up},j}\right),
\end{align}
which leads to the entanglement entropy $S(\hat{\rho}_\text{up})=N$.
Since the entanglement entropy increases linearly, an MPS simulation is inefficient.

For the second case in the main text, we consider an input state, where $N_1=N$, and $N_j=0$ for $2\leq j \leq M$.
In this case, from the analysis above, the reduced density matrix of the output state is written as
\begin{align}
    \hat{\rho}_\text{up}\approx \sum_{k=0}^{N}\frac{1}{2^{N}} \binom{N}{k}|k\rangle\langle k|_{\text{up},1},
\end{align}
and the entanglement entropy is given by
\begin{align}
    S(\hat{\rho}_\text{up})\approx \frac{1}{2}\log_2\left(\frac{\pi e N}{2}\right).
\end{align}
Since the entanglement entropy increases logarithmically, its MPS simulation can be efficiently performed.
Particularly, using $\chi=N+1$, the time complexity of a MPS simulation is $O(DMd^3\chi^3)=O(M^2 (N+1)^6)$.

In the case of lossy standard boson sampling, we can write the quantum state as
\begin{align}
    \hat{\rho}_\text{out}&=\prod_{j=1}^N \left[\mu \cos^2\theta_j|10\rangle\langle 10|_j+\mu \sin^2\theta_j|01\rangle\langle01|_j+\mu \sin\theta_j\cos\theta_j(|10\rangle\langle 01|_j+|01\rangle\langle 10|_j)+(1-\mu)|00\rangle\langle 00|_j\right] \\ 
    \to|\hat{\rho}\rangle\rangle&= \prod_{j=1}^N \left[\mu \cos^2\theta_j|30\rangle\rangle_j+\mu \sin^2\theta_j|03\rangle\rangle_j+\mu \sin\theta_j\cos\theta_j(|21\rangle\rangle_j+|12\rangle\rangle_j)+(1-\mu)|00\rangle\rangle_j\right].
\end{align}
Here, the index $j$ represents $\text{up},j$ and $\text{down},j$ in order, and for the vectorization, we merged the indices on each party as $|0\rangle\langle 0|\to |0\rangle\rangle$, $|0\rangle\langle 1|\to |1\rangle\rangle$, $|1\rangle\langle 0|\to |2\rangle\rangle$, and $|1\rangle\langle 1|\to |3\rangle\rangle$.
To obtain the matrix product operator (MPO) entanglement entropy (EE), we find the reduced density matrix for the vectorized state,
\begin{align}\label{eq:reduced}
    |\hat{\rho}\rangle\rangle\langle\langle \hat{\rho}|_\text{up}= \prod_{j=1}^N [(\mu \cos^2\theta_j|3\rangle\rangle&+(1-\mu)|0\rangle\rangle)(\mu \cos^2\theta_j\langle\langle3|+(1-\mu)\langle\langle0|)_{\text{up},j}+\mu^2 \sin^4\theta_j|0\rangle\rangle\langle\langle 0|_{\text{up},j} \nonumber \\
    &+\mu^2 \sin^2\theta_j\cos^2\theta_j(|2\rangle\rangle\langle\langle 2|_{\text{up},j}+|1\rangle\rangle\langle\langle 1|_{\text{up},j})].
\end{align}
Since MPO EE is additive, it is straightforward to obtain MPO EE.
For Figure \ref{fig:mpo}, we first generate a global Haar-random unitary matrix and find $\theta_j$ corresponding to the matrix.
We then use Eq.~\eqref{eq:reduced} to compute the MPO EE.
Especially when $\mu=\beta N^\gamma/N$ ($N_\text{out}=\beta N^\gamma$), assuming $M\geq N^2$ (collision-free) and an asymptotic limit $N\gg 1$, the average MPO EE for a bipartition $[1\cdots M/2]:[M/2\cdots M]$ can be approximated as
\begin{align}
    S_\alpha^{M/2}(|\hat{\rho}\rangle\rangle)&=O(N^{1-2(1-\gamma)\alpha}) ~~~~ \text{when}~~~~ \alpha\neq 1, \\ 
    S_1^{M/2}(|\hat{\rho}\rangle\rangle)&=O(N^{2\gamma-1}\log_2{N}).
\end{align}

\end{widetext}

\bibliography{reference.bib}

\end{document}